\documentclass[aps,prb,twocolumn,showpacs,floatfix,superscriptaddress,floatfix]{revtex4}
\usepackage[english]{babel}
\usepackage{graphicx}
\usepackage{color}
\usepackage[utf8]{inputenc}
\usepackage{pstricks,pst-grad,color}
\usepackage{graphicx,amssymb}
\usepackage{amsmath}
\usepackage{amssymb}

\begin{document}

\title{Coexistence of light and heavy surface states in a topological multi-band Kondo insulator }

\author{Robert Peters}
\email[]{robert.peters@riken.jp}
\affiliation{Department of Physics, Kyoto University, Kyoto 606-8502,  Japan}

\author{Tsuneya Yoshida}
\affiliation{RIKEN, Condensed Matter Theory Laboratory,  Wako, Saitama 351-0198, Japan}

\author{Hirofumi Sakakibara}
\affiliation{Computational Condensed Matter Physics Laboratory, RIKEN,
  Wako, Saitama 351-0198, Japan}

\author{Norio Kawakami}
\affiliation{Department of Physics, Kyoto University, Kyoto 606-8502,  Japan}

\date{\today}

\begin{abstract}
We analyze the impact of strong correlations on the
surface states of a three-dimensional topological Kondo insulator.
An important observation is that correlations are strongly increased
at the surface, which leads to an energetical confinement of
the surface $f$-electrons into a narrow window around the Fermi energy at low enough temperature. This 
correlation effect in combination with the nontrivial topology has two remarkable consequences: (i) coexistence of light and heavy surface states at low temperatures. While heavy surface
states are formed directly in the surface layer, light surface
states are formed in the next-nearest surface layer. Furthermore, (ii)
with increasing the temperature, the heavy surface states become
incoherent and only light surface states can be observed. The
coexistence of light and heavy surface states is thereby a
remarkable characteristic of the combination of strong correlations
and nontrivial topology. We believe that these results are applicable to the candidate topological Kondo insulator SmB$_6$.
\end{abstract}

\maketitle

\section{Introduction}
Strongly correlated topologically nontrivial materials are a class of
materials, which combines two fields of condensed matter physics recently
attracting enormous interest. On the one hand, strong correlations,
which are often found in $d$- and $f$-electron systems, are the origin
of intriguing phenomena such as
heavy-quasi-particles, magnetism, unconventional superconductivity, etc.
On the other hand, nontrivial topology leads to the emergence of
symmetry protected surface states \cite{RevModPhys.82.3045,RevModPhys.83.1057}.
Combining both fields results in highly interesting and
 nontrivial systems and raises questions about correlation
effects and ordering of the topologically protected surface states \cite{Pesin2010,PhysRevLett.106.100403,PhysRevLett.107.010401,0953-8984-25-14-143201,PhysRevB.85.125113}.

Candidates for such strongly correlated and topologically nontrivial materials are for example $f$-electron systems with strong spin-orbit interaction which are insulating, such as SmB$_6$ \cite{Fisk1995798,coleman2007,Takimoto2011} or CeOs$_4$As$_{12}$ \cite{PhysRevB.85.165125}.
SmB$_6$ is a long-known Kondo insulator. Strongly interacting
$f$-electrons hybridize with conduction 
$d$-electrons in a way that a bulk gap opens at the Fermi energy,
which is experimentally confirmed by a large increase of the
resistivity below $40$K \cite{PhysRevLett.22.295}. However, a puzzle
has remained since the discovery
of this Kondo insulator, namely the saturation of the resistivity
below $4$K \cite{PhysRevB.20.4807}. It has turned out that this saturation can be
attributed to quite robust surface states
\cite{PhysRevB.88.180405,PhysRevX.3.011011,Kim2013,Li05122014,Kim2014}. The
existence of these robust surface states led to the proposal that this
material might be topologically nontrivial
\cite{Takimoto2011,PhysRevLett.104.106408,PhysRevB.85.045130,PhysRevLett.111.226403},
 making SmB$_6$ a topological Kondo insulator, which
  has led to extensive theoretical research on this topic
  \cite{PhysRevLett.104.106408,PhysRevB.85.045130,PhysRevLett.111.226403,PhysRevLett.110.096401,PhysRevB.85.125128,PhysRevB.88.035113,PhysRevB.89.085110,PhysRevB.89.245119,PhysRevB.90.081113,PhysRevB.90.195144,PhysRevB.90.235107,PhysRevB.90.165127,PhysRevB.90.155314,PhysRevB.90.115109,PhysRevB.90.201106,PhysRevB.91.245127,PhysRevLett.112.226402,PhysRevB.87.165109,yoshida2015}.
Indeed, first principles calculations
seem to confirm this prediction, demonstrating that SmB$_6$ is a
strong topological insulator with topologically protected surface states
at $\Gamma$- and $X$-points \cite{Takimoto2011,PhysRevLett.110.096401},
which have also 
been found experimentally by ARPES \cite{Jiang2013,Neupane2013,PhysRevB.88.121102,PhysRevLett.111.216402,PhysRevX.3.041024,Xu2014}.
Thus, SmB$_6$ might be the model system
for a strongly correlated topological insulator.

The aim of this paper is to study the interplay of strong correlations and nontrivial topology in 3D $f$-electron materials such as SmB$_6$. 
We will therefore use a model Hamiltonian exhibiting a simplified version of the band structure of SmB$_6$, including all essential orbitals to describe a topological Kondo insulator with metallic surface states at the $\Gamma$- and $X$-point in the Brillouin zone. This model will be too simple to explain every detail of SmB$_6$, but it is a realistic starting point to analyze the interplay between strong correlations and nontrivial topology.
In this paper we focus especially on the impact of the correlations on the topological surface states.
 Related to this topic is an open question on the group velocities of the surface states in SmB$_6$: While theoretical calculations predict heavy surface states, most experiments observe rather light ones. Recent transport measurements at very low temperature, on the other hand, indicate the existence of heavy surface states. Alexandrov {\it et al.} \cite{PhysRevLett.114.177202}   proposed a solution to the problem why the surface states should be light. They assumed layer-dependent Kondo temperatures and performed model calculations based on the slave-boson mean field theory. If the Kondo temperature at the surface is much lower than in the bulk, then there would be light surface states, if the temperature in the experiments is larger than the surface Kondo temperature.

We here use dynamical mean field calculations\cite{Georges1996} to analyze the topological surface states in a model Hamiltonian describing a strongly correlated topologically nontrivial Kondo insulator.
We demonstrate that
due to a significant increase of correlations directly at
the surface, the spectrum at $T=0$K is composed of
light and heavy surface states. 
We show that heavy surface states emerge in the outermost surface
layer, where the $f$-electrons are energetically confined close to
the Fermi energy.  However, the nontrivial topology of the system
ensures the existence of surface states connecting the bulk states
below the Fermi energy with the states above the Fermi energy, which
cannot be accomplished by the outermost surface layer alone.
Due to this interplay between topology and strong correlations, light
``surface'' bands emerge in the next-nearest-surface layer slightly away from the Fermi energy.
We furthermore observe in our model that similar to the calculations  by Alexandrov {\it et al.} \cite{PhysRevLett.114.177202} the heavy surface states vanish when the temperature is increased so that only light surface states exist.

The remainder of this  paper is organized as follows: In the next section we explain the model and the methods that we use in our calculations. This is followed by a section analyzing the topology of the interacting model. Thereafter, we show results for the momentum-resolved and local density of states elucidating the correlation effects on the topological surface states.

\section{Model and method}
                Motivated by SmB$_6$, which is a good candidate for a strongly correlated and topologically nontrivial material, we use a simplified band structure of SmB$_6$ as the noninteracting part of our Hamiltonian. The band structure is based on a first principles calculation using the WIEN2k package.\cite{Wien2K,Kunes20101888,PhysRevB.56.12847,PhysRevB.65.035109} Being interested in a simple model Hamiltonian, which nevertheless can describe the essential properties due to the interplay of strong correlations and nontrivial topology, we only use the $e_g$-orbitals of the $d$-electrons and the $\Gamma_8$-quartet of the $f$-electrons in our study. Although this simplified band structure will not be able to describe every detail of the low temperature physics of SmB$_6$, it provides us with generic properties inherent in three-dimensional strong topological Kondo insulators.

\begin{figure}[t]
\begin{center}
\includegraphics[width=0.48\linewidth]{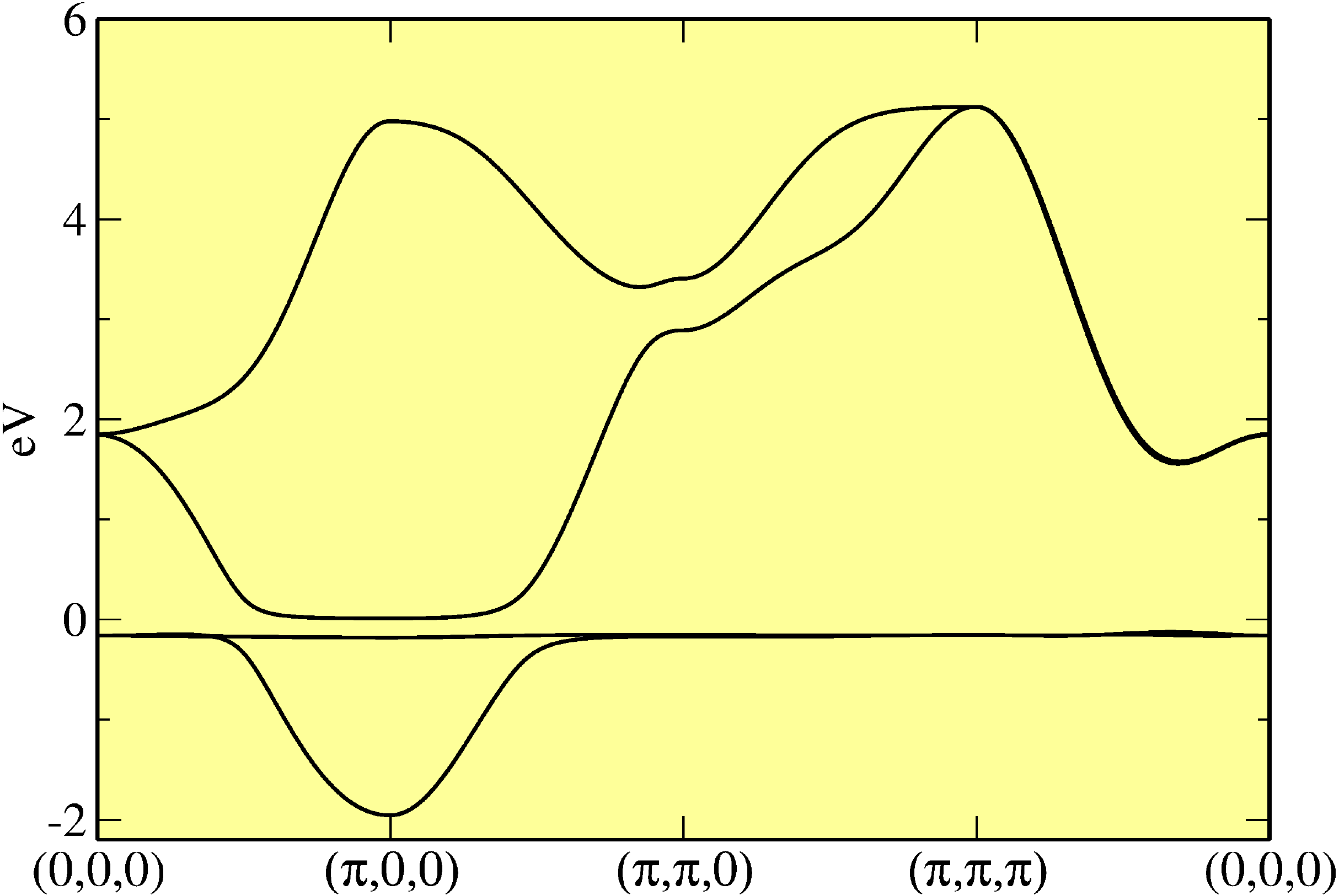}
\includegraphics[width=0.48\linewidth]{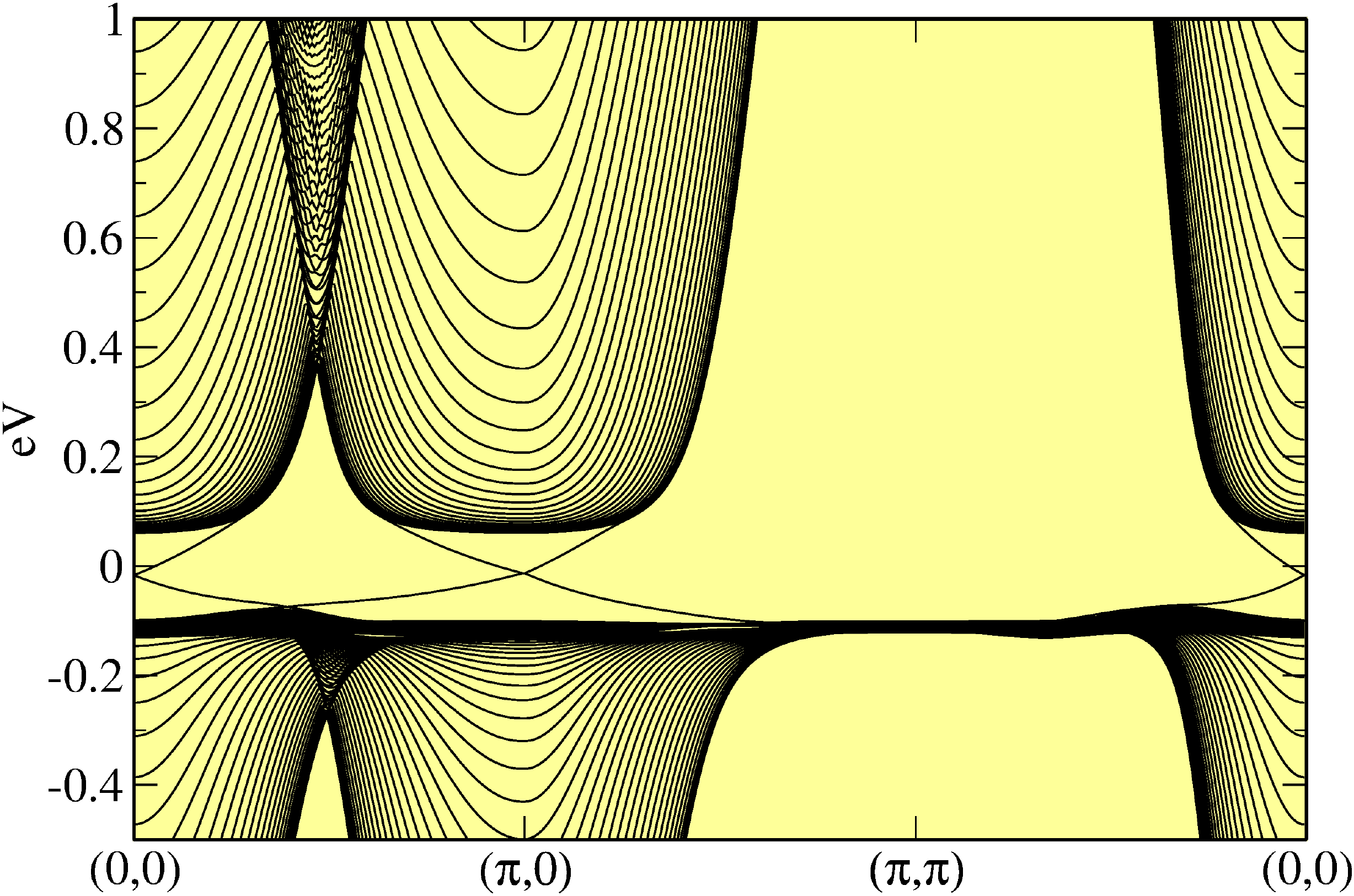}
\end{center}
\caption{(Color online) Left: Bulk band-structure showing the wide $d$-bands and  narrow $f$ bands.  $d$- and $f$-bands are hybridized and form a gap at the Fermi energy. Right: Magnification around the Fermi energy of a  
  noninteracting slab calculation with open  boundaries in
  $z$-direction. Clearly visible is the bulk gap and the topological
  surface states at $\Gamma$- and $X$-points.  
\label{LDA}}
\end{figure}

Thus, our model consists of a 3D cubic lattice and includes $8$ orbitals per lattice site. This model conserves time-reversal- and inversion symmetry and
exhibits topological surface states at the $\Gamma$- and the $X$-points in the Brillouin zone similar to SmB$_6$.
The spin orbit interaction is fully included. The band width of the $d$-electrons in our model, which is approximately $7$eV, agrees with that of SmB$_6$  and sets the energy scale through out this paper. The band width of the noninteracting $f$-electrons is $0.2$eV.
We note that presenting the full Hamiltonian in our effective tight binding model should include several hundreds of hopping terms, which are difficult to accommodate in the paper. Instead, we would like to mention the essential points due to the nontrivial band structure in Fig. \ref{LDA}, which are relevant to the Kondo insulator; Because of inversion of bands with different parity at $3$ time-reversal invariant points in the Brillouin zone, the noninteracting model is topologically nontrivial.\cite{PhysRevB.76.045302,PhysRevLett.98.106803,PhysRevLett.104.106408,PhysRevX.2.031008}
The noninteracting band
structure close to the 
Fermi energy with open boundaries in $z$-direction, shown in the
Fig. \ref{LDA} (right), demonstrates the existence of surface
states at $\Gamma$- and $X$-points in the Brillouin zone.

In order to study a Kondo insulator, we include a strong local interaction into the $f$-electron orbitals, which will lead to a Kondo effect at low temperatures. The inclusion of 4 $f$-electron bands and especially the coupling between these bands leads to a strong increase of correlations in the model compared to single $f$-electron band models.
We use the following multi-band Hubbard Hamiltonian for the  $f$-electron $\Gamma_8$-bands
\begin{eqnarray*}
H_{\Gamma_8}&=&U\left(n_{\Gamma_{8,a\uparrow}}n_{\Gamma_{8,a\downarrow}}+n_{\Gamma_{8,b\uparrow}}n_{\Gamma_{8,b\downarrow}}\right)\\
&&+U^\prime(n_{\Gamma_{8,a\uparrow}}+n_{\Gamma_{8,a\downarrow}})(n_{\Gamma_{8,b\uparrow}}+n_{\Gamma_{8,b\downarrow}})\nonumber\\
&&+J\vec{S}_{\Gamma_{8,a}}\cdot\vec{S}_{\Gamma_{8,b}},
\end{eqnarray*}
where $n_{\Gamma_{8,\{a,b\}\sigma}}$ is the occupation operator for orbital
$\Gamma_{8,\{a,b\}}$ with
pseudo-spin direction $\sigma=\{\uparrow,\downarrow\}$. Finally, $\vec{S}_{\Gamma_{8,\{a,b\}}}$ corresponds to the spin operator for
the $\Gamma_{8,a}$- and $\Gamma_{8,b}$-orbitals. We use the following interaction
strengths $U=4$eV, $U^\prime=2$eV and $J=1$eV, but have confirmed
that our results do qualitatively not depend on these interaction
strengths at $T=0$. 
In this study we fix the chemical potential of the $f$-electrons in a way that these bands are occupied by $n_f=3$ electrons and leave the analysis of different electron numbers as a future study. For this filling the $f$-electron bands are more than half-filled, which qualitatively agrees with the situation of SmB$_6$.

We use the real-space dynamical mean field theory (DMFT) for a system of $20$ layers, where each layer of the system is mapped
onto its own impurity model, which is solved self-consistently.
Nonlocal correlations are thereby excluded and all self-energies are local. However, local fluctuations are fully included into this  study.
This usage of real space DMFT enables us to study layer-dependent
correlation effects and the impact of these
correlations on the surface states. 
Because the lattice has translational symmetry in the $x$-$y$-plane, we can impose periodic boundary conditions within each layer. However, due to the open boundaries in $z$-direction, the effective impurity models depend, via the bath Green's function, on the layer.
We have recently used similar methods to analyze correlation effects in the topologically nontrivial Hubbard model \cite{PhysRevB.85.165138} and $f$-electron superlattices \cite{PhysRevB.88.155134}. 
We use the numerical renormalization group (NRG)\cite{wilson1975,Bulla2008} to solve the
resulting impurity models, which is able to calculate
real-frequency spectral function for arbitrary temperatures and
resolve even small structure at the Fermi energy
\cite{Peters2006,Weichselbaum2007}. 
By using the
NRG as impurity solver we can provide real-frequency self-energies and
Green's function without the need of an analytic continuation of
imaginary-time data, and we are able to
analyze details in the band structure from $6$eV (roughly the band
width of the $d$ electrons) to $0.0001$eV (approximately the Kondo
temperature of the surface layer in our calculations).

\section{Topological structure} 
Because of inversion of bands with different parity at $3$ time-reversal invariant points in the Brillouin zone, the ground state of the noninteracting model is topologically nontrivial.
\begin{figure}[t]
\begin{center}
\includegraphics[width=\linewidth]{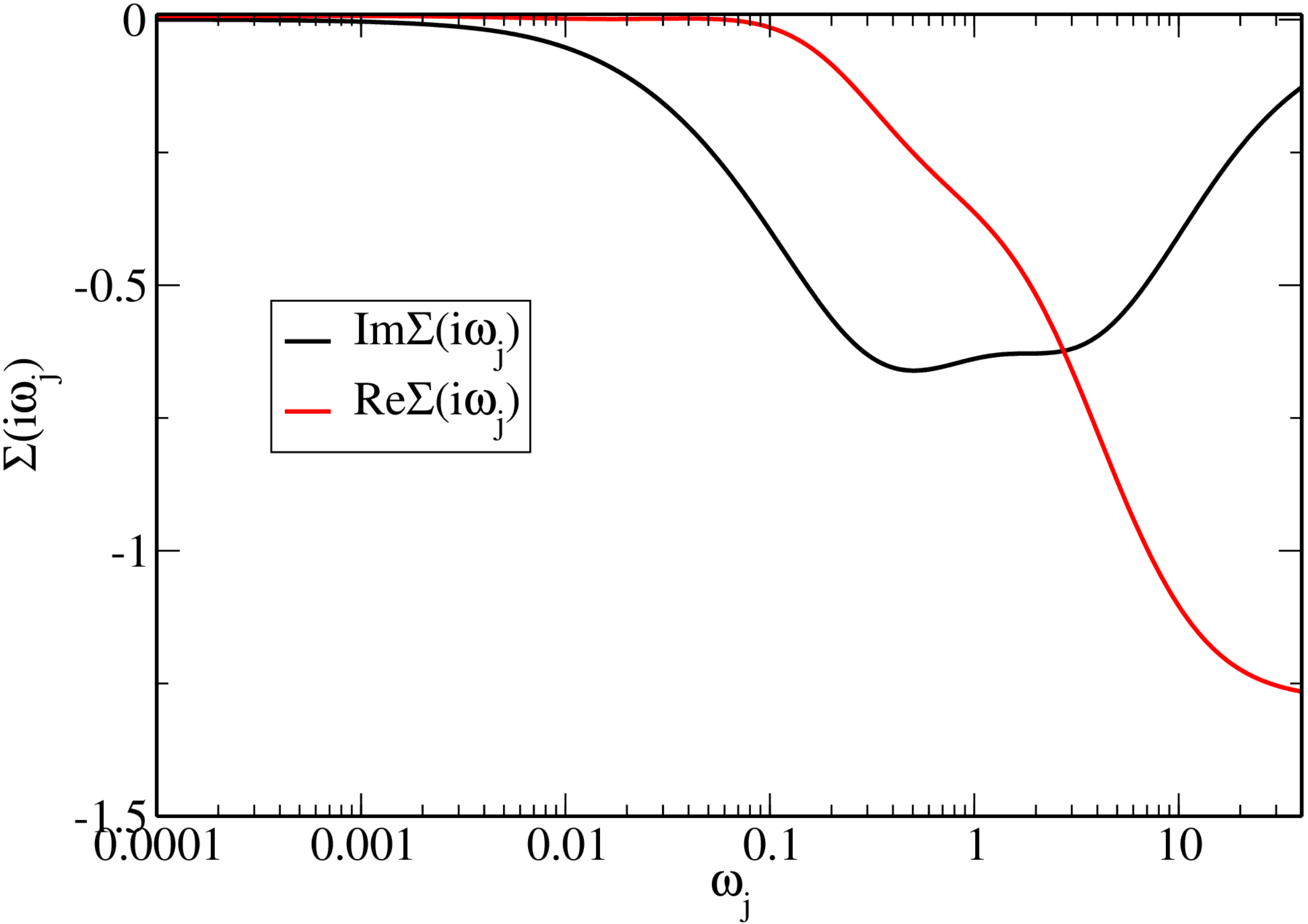}
\end{center}
\caption{(Color online) Bulk self-energy in Matsubara
  frequencies for the 
   $f$-electron bands for $T=10^{-7}$eV, which is much
  smaller than the coherence temperature of the surface layer.
\label{self_matsubara}}
\end{figure}
Next, we confirm that even in the presence of correlations this model is a strong topological insulator.
In the presence of electron correlations, the $Z_2$ invariant for
strong topological insulators is defined in terms of the
single-particle Green's function, $\hat{G}(i\omega,\vec{k})$. 
In Ref. \cite{PhysRevX.2.031008}, it has been pointed out that the
topological invariant for correlated systems can be obtained from the
topological Hamiltonian
$\hat{h}'(\vec{k}):=\hat{h}(\vec{k})+\hat{\Sigma}(i\omega=0,\vec{k})=-\hat{G}^{-1}(0,\vec{k})$, if
the single-particle Green's function $\hat{G}(i\omega,\vec{k})$ is
nonsingular; $\mathrm{det}\hat{G}(i\omega,\vec{k})\neq0$ and
$\mathrm{det}\hat{G}^{-1}(i\omega,\vec{k})\neq0$.  
$\hat{h}(\vec{k})$ is thereby the Fourier transformation of the
hopping Hamiltonian and $\hat{\Sigma}$ is the self-energy matrix,
which has only nonvanishing entries at the diagonal elements of the $f$-orbitals.
Then the Green's function $\hat{G}(i\omega,\vec{k})$ can be
continuously deformed without any singularity to
$\hat{g}(i\omega,\vec{k}):=[i\omega\mathbb{I}+\hat{G}^{-1}(0,\vec{k})]^{-1}$ by using
$\hat{G}(i\omega,\vec{k},\lambda):=\lambda \hat{G}(i\omega,\vec{k}) +(1-\lambda)
\hat{g}(i\omega,\vec{k})$ with $\lambda\in[0,1]$. 

Here we show that $\hat{G}^{-1}(i\omega,\vec{k})$ is nonsingular and thus,
we can use the topological Hamiltonian for the calculation of the $Z_2$
invariant. 
First we show $\mathrm{det}\hat{G}(i\omega,\vec{k})\neq 0$.
Since the band width is finite, the only possible way for
$\mathrm{det}\hat{G}(i\omega,\vec{k})=0$ is a divergence of the self-energy $\hat{\Sigma}(i\omega,\vec{k})$.
Because of the DMFT approximation, the self-energy is momentum
independent. Furthermore, within the current model off-diagonal
elements of the self-energy between different orbitals vanish.
The diagonal element, $\Sigma(i\omega)$, is shown in
Fig.~\ref{self_matsubara}, and does not exhibit any divergence. Therefore,
$\mathrm{det}\hat{G}(i\omega,\vec{k})\neq 0$ holds for any $\vec{k}$ and
$i\omega$.  
Because $\mathrm{det}\hat{G}^{-1}(i\omega,\vec{k})= 0$ corresponds to the
closing of the bulk gap, we can also rule out this possibility,
because a bulk gap is observed in our calculations.

Thus, we can obtain the $Z_2$ invariant for the correlated system from
the topological Hamiltonian $\hat{h}'(\vec{k})$.  
We note that our model is inversion symmetric and the following
equation holds
\begin{eqnarray*}
\hat{P}\hat{h}'(\vec{k})\hat{P}^{-1}&=&\hat{h}'(-\vec{k}), \\
\hat{P}&=&\begin{pmatrix}
\begin{pmatrix}1&\ldots&0\\
 &\ddots&\\
0&\ldots&1
\end{pmatrix}_{e_g}  &0\\
0&\begin{pmatrix}-1&\ldots&0\\
 &\ddots&\\
0&\ldots&-1
\end{pmatrix}_{f}
\end{pmatrix},
\end{eqnarray*}
because the hybridization between $d$- and $f$-orbitals is given by an
odd function in the momentum space.

Thus, we can apply the simplified formula for inversion symmetric
systems \cite{PhysRevB.76.045302,PhysRevLett.98.106803} and calculate
the $Z_2$ invariant via
\begin{eqnarray}
(-1)^{\nu}&=&\Pi_{\vec{k}^*_i,\alpha \in occ.}\delta_{\vec{k}^*_i,\alpha},
\label{eq: Z2}
\end{eqnarray}
where $\vec{k}^*_i$ are the time-reversal invariant momenta [i.e.,
  $(0,0,0)$, $(\pi,0,0)$, $(0,\pi,0)$, $(0,0,\pi)$, $(\pi,\pi,0)$,
  $(\pi,0,\pi)$,  $(0,\pi,\pi)$, $(\pi,\pi,\pi)$ ] and 
$\delta_{\vec{k}^*_i,\alpha}$ denote the eigenvalues of the occupied
orbitals  $\alpha$ of the inversion operator $\hat{P}$ at $\vec{k}^*_i$.

Taking into account the Kramer's degeneracy of the system, we find that
$\delta_{\vec{k}^*_i,\alpha}$ becomes $-1$ at  $(\pi,0,0)$,
$(0,\pi,0)$, $(0,0,\pi)$ and $+1$ at all $5$ other time-reversal invariant
points.
Therefore, from Eq. ($\ref{eq: Z2}$), we conclude that our model is a strong
topological insulator \cite{Takimoto2011,PhysRevLett.110.096401}.

\section{Results}
\begin{figure}[t]
\begin{center}
\includegraphics[width=\linewidth]{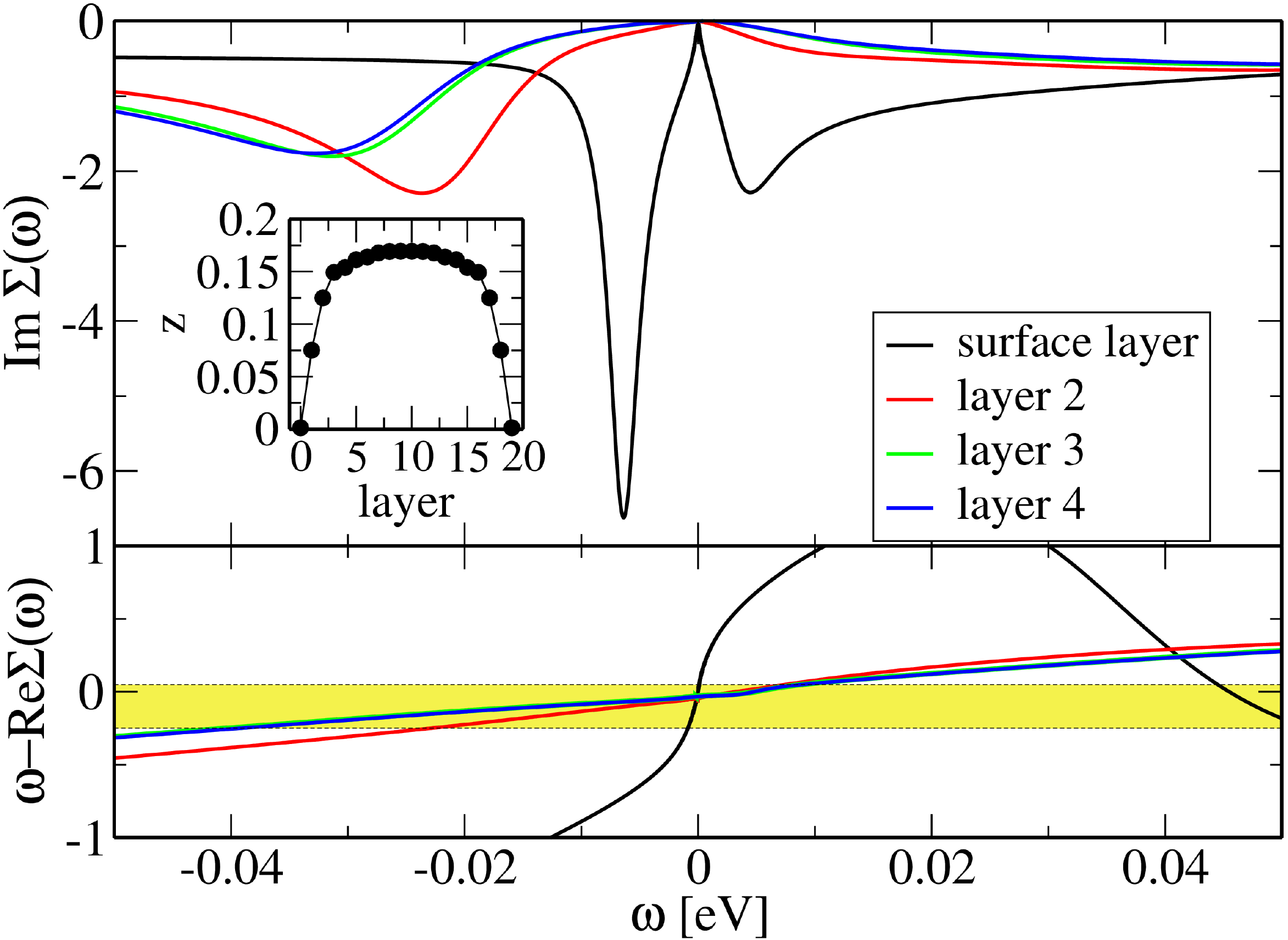}
\end{center}
\caption{(Color online) Calculated self-energies of the $f$-electron
  quartet. Top: Imaginary part of the self-energy for different
  layers. The inset shows the layer dependence of the
  quasi-particle weight   $z=1/(1-\partial
  \text{Re}\Sigma(\omega)/\partial\omega)$ for a slab calculation
  using $20$ layers. Although the quasi-particle weight is very small
  at the surface, it does not vanish.
Bottom: Real part of the self-energy for
  different layers. The yellow marked area corresponds to the energies
  of the noninteracting $f$-electrons.
\label{self}}
\end{figure}

We show real-frequency self-energies for several layers in Fig. \ref{self}.
The used two-particle interactions lead to two types of excitations in
the $f$-orbitals:
an intra-band particle-hole 
excitation
(not shown in Fig. \ref{self}) and the
Kondo effect close to the Fermi energy. Exactly at the Fermi energy, the
imaginary part of the 
self-energy vanishes at $T=0$K, which gives rise to a highly correlated Fermi
liquid, characteristic for heavy fermion systems.
The self-energy around the Fermi energy is thereby dominated by
inter-band correlations in the $f$-electron bands.
As it is visible from the behavior of the self-energy around the Fermi
energy, $\omega=0$, and also
from the quasi-particle weight  $z=1/(1-\partial
\text{Re}\Sigma(\omega)/\partial\omega)$, the surface layer is much
more strongly correlated than all other layers.
This increase in the correlation can be attributed to the open surface
which leads to a change in the hybridization between $f$-electrons and
$d$-electrons \cite{PhysRevLett.114.177202}. 
The large slope in the real part of the self-energy of the surface
layer has thereby a very important consequence: $f$-electron bands in
the surface layer can only emerge in a very narrow energy region
around the Fermi energy, namely when the equation
$\omega-\text{Re}\Sigma(\omega)-\epsilon_{\vec{k}}-\mu=0$ can be
fulfilled. In the lower panel of Fig. \ref{self} we have marked the
energy region of the noninteracting $f$-electron bands yellow. 
Only for $\omega$ for which $\omega-\text{Re}\Sigma(\omega)$ is within the
marked area, bands can emerge in the interacting model.

\begin{figure*}
\includegraphics[width=0.49\linewidth]{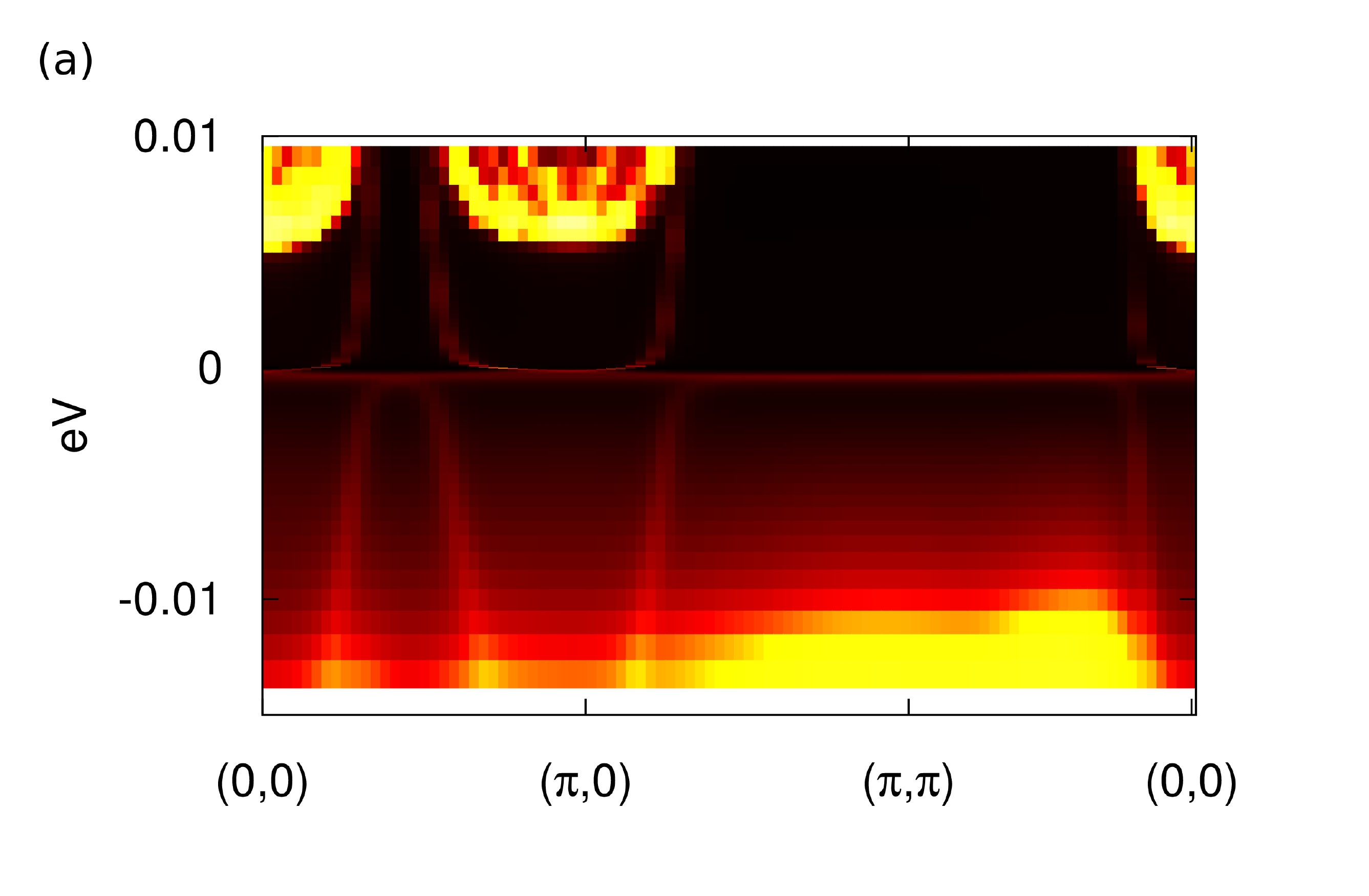}
\includegraphics[width=0.49\linewidth]{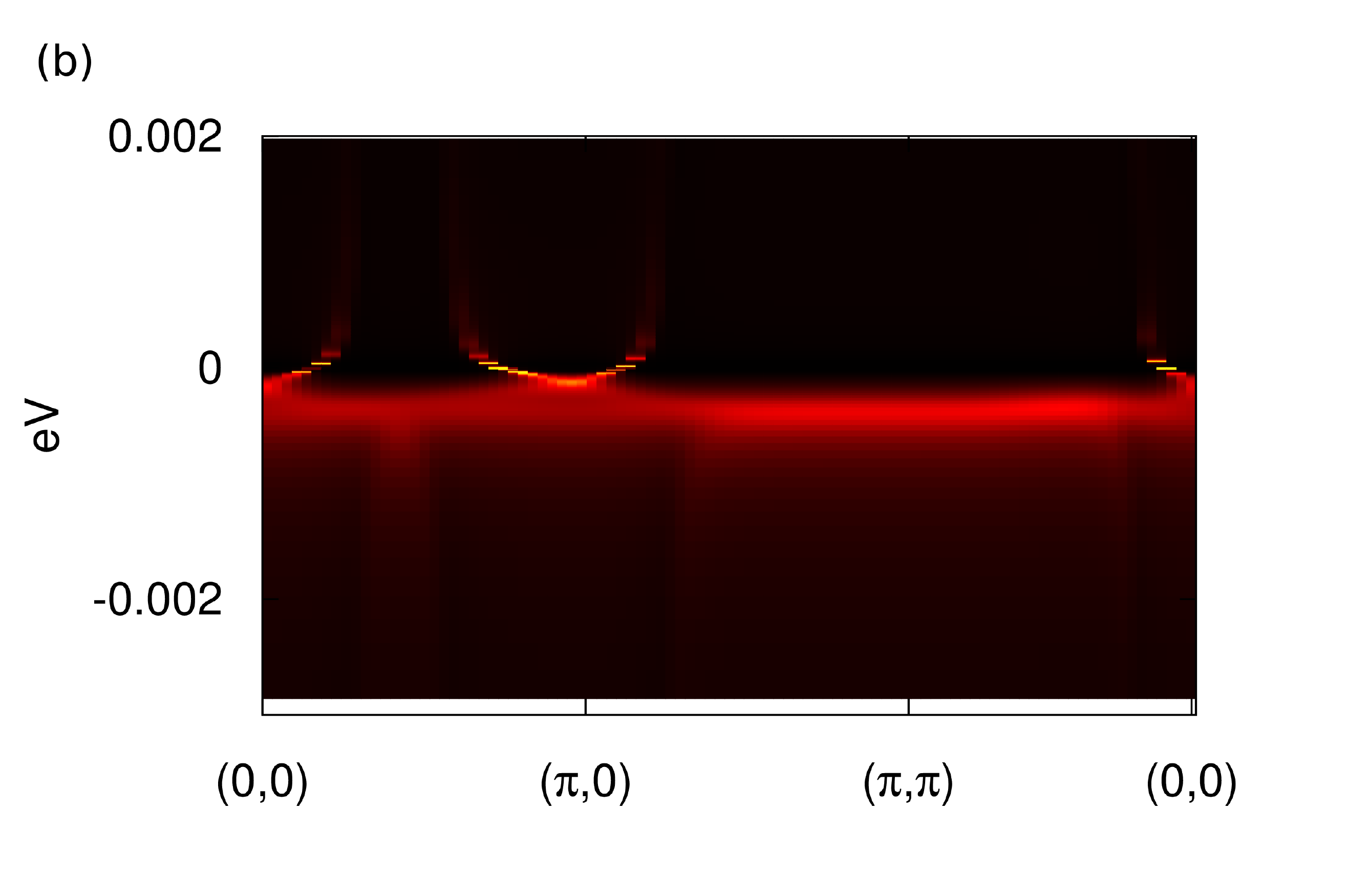}
\includegraphics[width=0.49\linewidth]{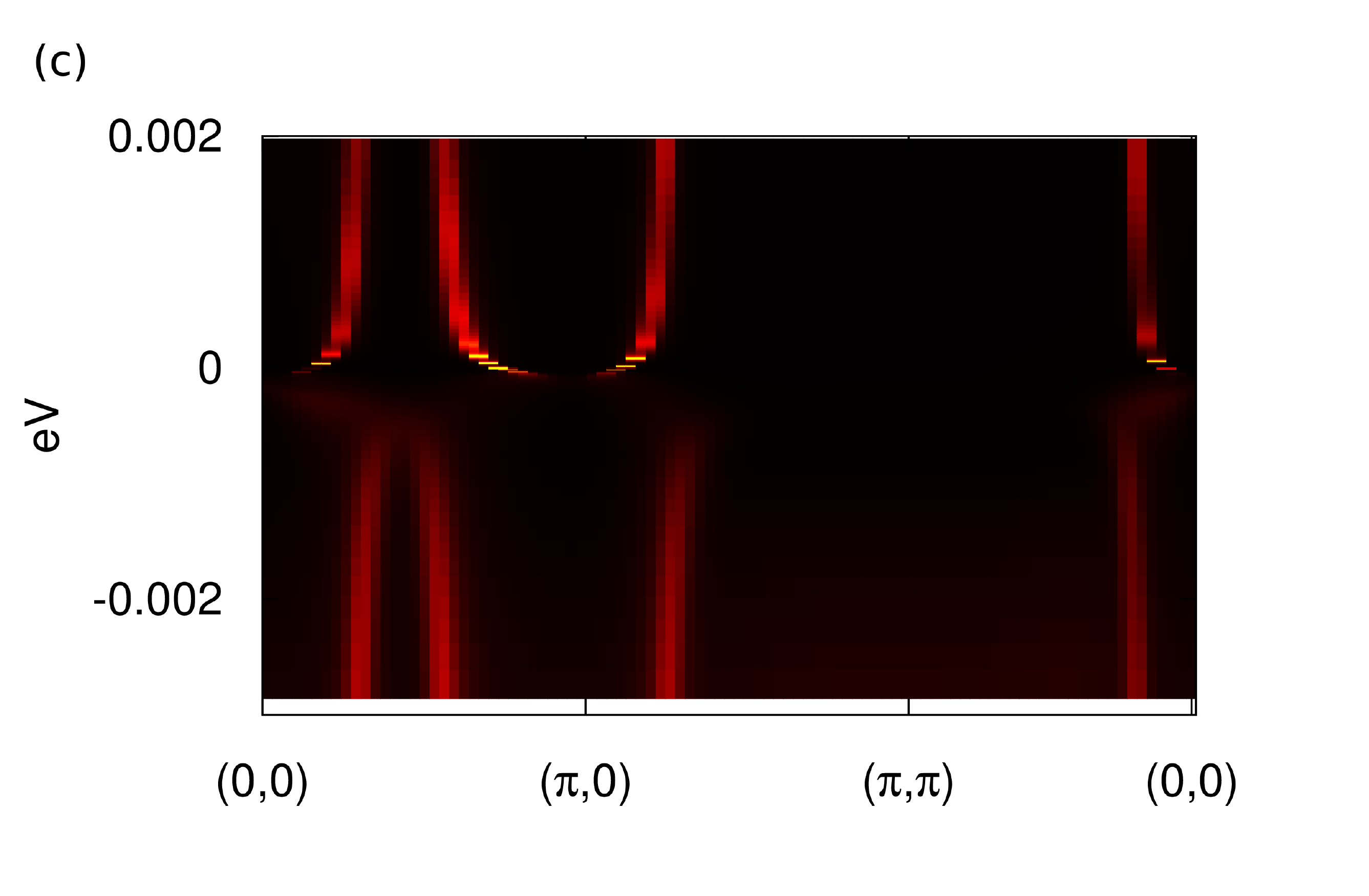}
\includegraphics[width=0.49\linewidth]{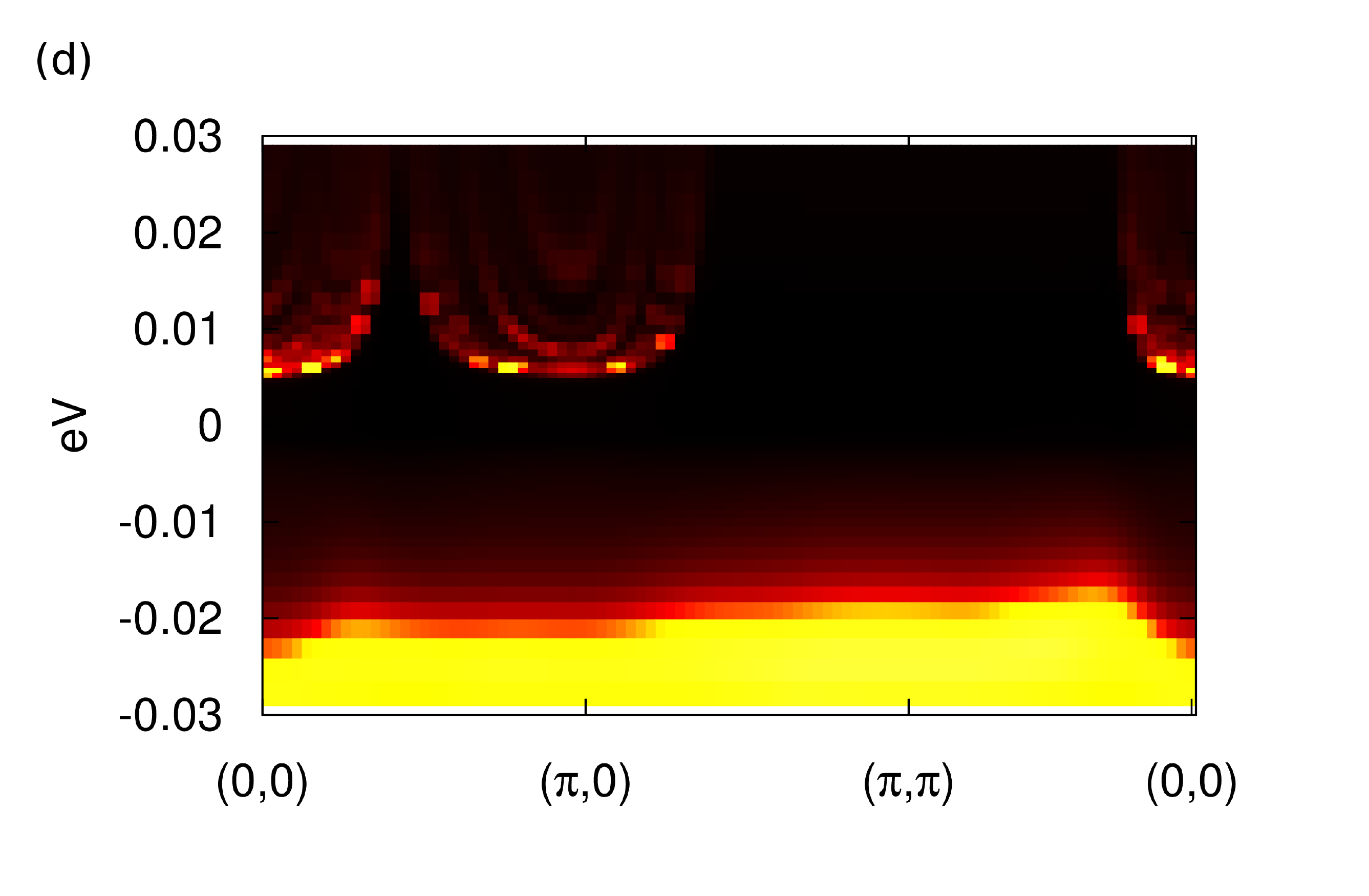}
\caption{(Color online) Momentum resolved spectral functions at $T=0$ around the
  Fermi energy. (a) Spectrum of all layers and all orbitals; (b) Spectrum of the surface $f$ electrons; (c)
  Spectrum of the $f$-electrons of the next-nearest surface layer;(d) Spectrum of the $f$ electrons in a bulk layer. Notice the different energy scales for each panel. In these color-plots black corresponds to weak, red to
  intermediate, and yellow to strong intensity.
\label{in_gap}}
\end{figure*}

Let us now analyze the impact of these correlations on the topological
surface states.
In Fig. \ref{in_gap}, we show momentum resolved Green's functions
for model calculations consisting of $20$ layers with open boundaries in $z$-direction.
The inclusion of correlation effects leads to a reduction of the bulk gap width from $\Delta\approx 0.1$eV in the noninteracting model to $\Delta\approx 0.02$eV in the interacting model. However, a bulk gap still exists, see panel (d) in Fig. \ref{in_gap}.
Besides changing the width of the gap, the layer-dependent self-energy has strong influence on the
surface states.
In Fig. \ref{in_gap}(a) we show the whole spectrum of all layers and all orbitals.
We can clearly observe the presence of in-gap states in this plot, which emerge within the surface layers.
These surface states emerge as light bands from the bulk states below
the Fermi energy and evolve until close to the Fermi energy. At an
energy scale corresponding to the coherence temperature of the surface
layer the energy-momentum dispersion of these bands changes
dramatically, see panel (a). The light surface states
originating in the bulk states below the Fermi energy
 form a narrow and flat band
below the Fermi energy. From this flat surface band {\it
  heavy} Dirac cones are formed around the Fermi energy. For slightly
higher energies, these bands change again into
light surface states which are connected to the bulk states above the Fermi
energy.

{\it How can we understand the simultaneous existence of light and heavy surface states?} The topology of this Hamiltonian ensures the
existence of surface states which connect the bulk states below the
Fermi energy with the bulk states above the Fermi energy. However, the
self-energy of the outermost surface layer, shown in Fig. \ref{self},
effectively confines the $f$-electrons of this layer into a small
energy window around the Fermi energy. From the real-part of the self-energy we can read off this energy window  to be $[-8\cdot
  10^{-4}\text{eV},1\cdot 10^{-4}\text{eV}]$.
 In order to confirm this, we show in
panel (b) of Fig. \ref{in_gap} the spectral function of the 
$f$-electrons at the surface. Clear energy bands are only visible
within the above mentioned energy window and appear as heavy surface states.
However, the bulk gap is wider than this
energy region.
 For the energies lying between this
narrow window of the outermost surface layer and the bulk states, the
surface states ensured by the topology are formed in the next-nearest surface
layer, see panel (c). Because in this layer, the correlations
are not as strong as in the outermost surface layer, these surface
states appear as light surface states.

The electron number does only weakly depend on the layer.  Due to the Kondo effect, which drives the system towards an integer number of $f$-electrons, the difference between the bulk layer and the surface layer is very small: approximately $\Delta n\approx 0.001$ per orbital. Thus, the difference in the correlation strength cannot be explained by a change in the particle number. Furthermore, the effects of double counting due to the combination of an LDA band structure and DMFT, which are usually taken into account by the inclusion of Hartree terms proportional to the particle number, do not influence the layer dependence of the effective electron mass.
The necessary conditions to observe the described phenomena in a topological Kondo insulator are thereby the formation of a Fermi liquid state at low temperatures and a strong enhancement of the effective electron mass at the surface. Then, the coexistence of light and heavy surface states naturally follows at low temperatures, which does not depend on the detail of the band structure.

The emergence of heavy surface
states in the outermost layer and light surface states in the
next-nearest neighbor can only occur due to the layer-dependence of
the self-energy. We note here that the layer, in which the light surface states emerge, depend on how the strength of the correlations changes with the layer. In our model, correlations are strongly enhanced at the surface and are already much weaker in the next-nearest-surface layer, where the light states are observed. If the change of the correlation strength would occur inside the material, these light states will also appear  inside. Such a drop in the correlation strength not at the surface but within the solid could be seen e.g. in heterostructures of topologically nontrivial materials.

\begin{figure*}
\includegraphics[width=0.49\linewidth]{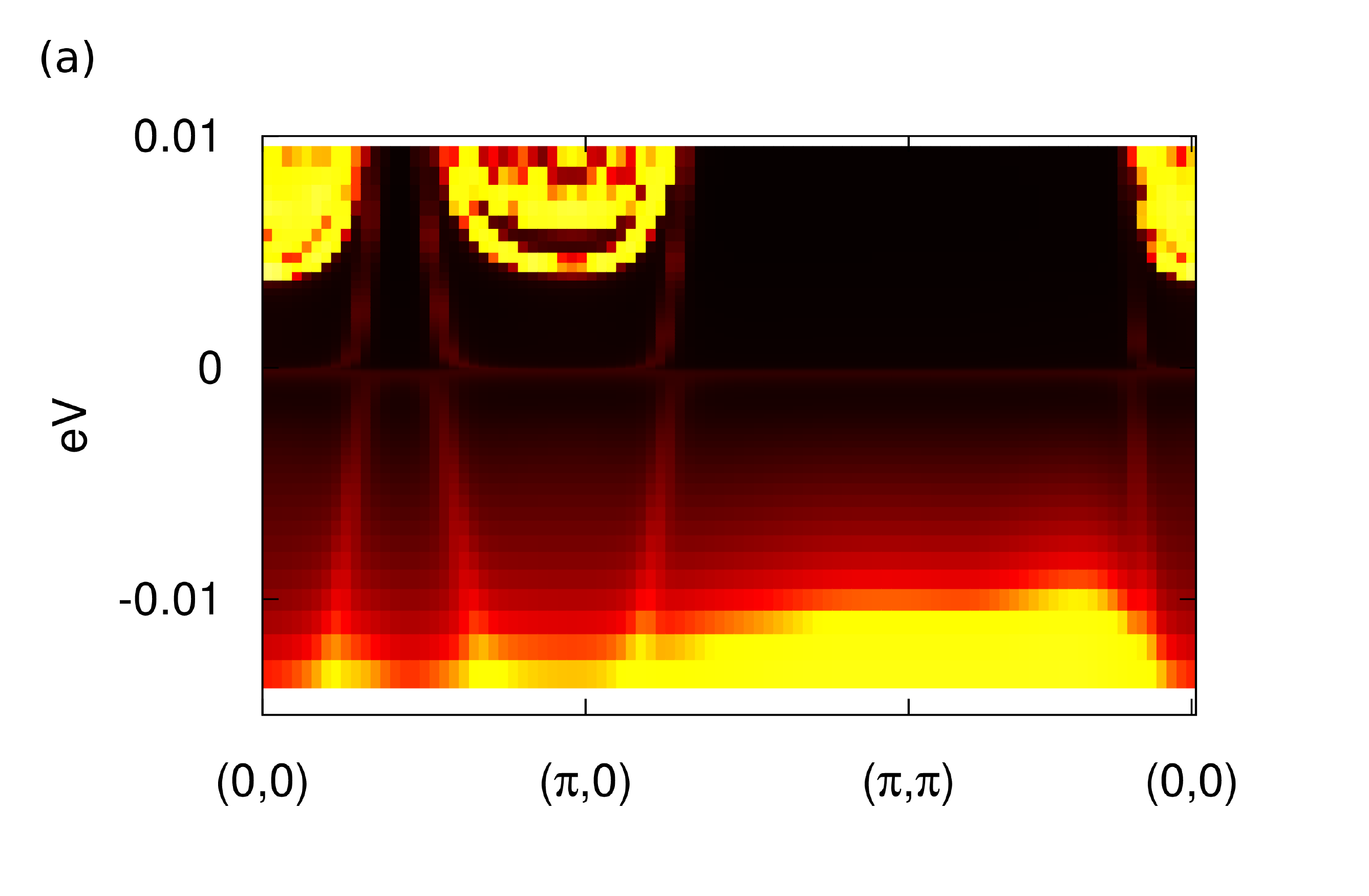}
\includegraphics[width=0.49\linewidth]{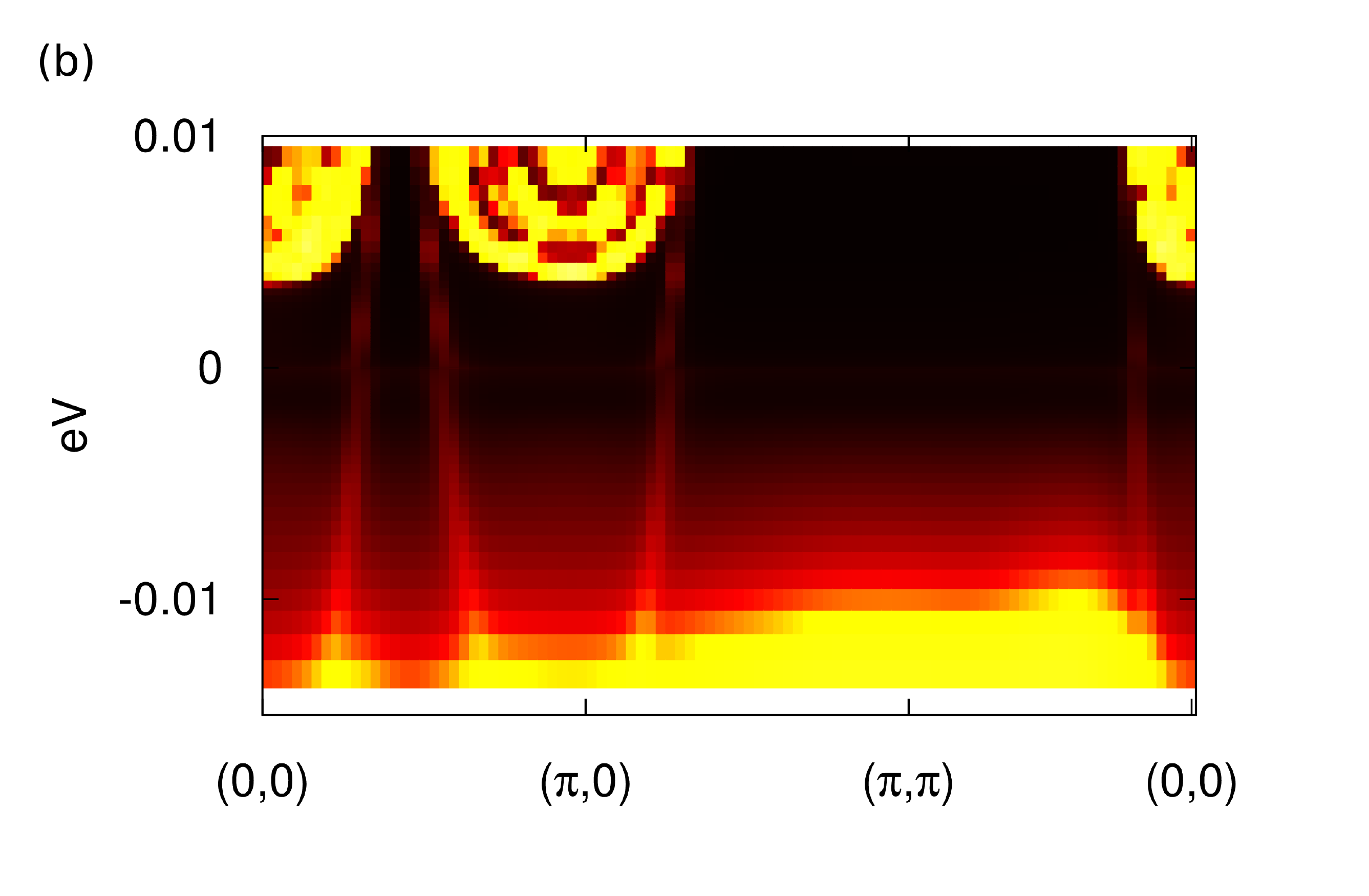}
\includegraphics[width=0.49\linewidth]{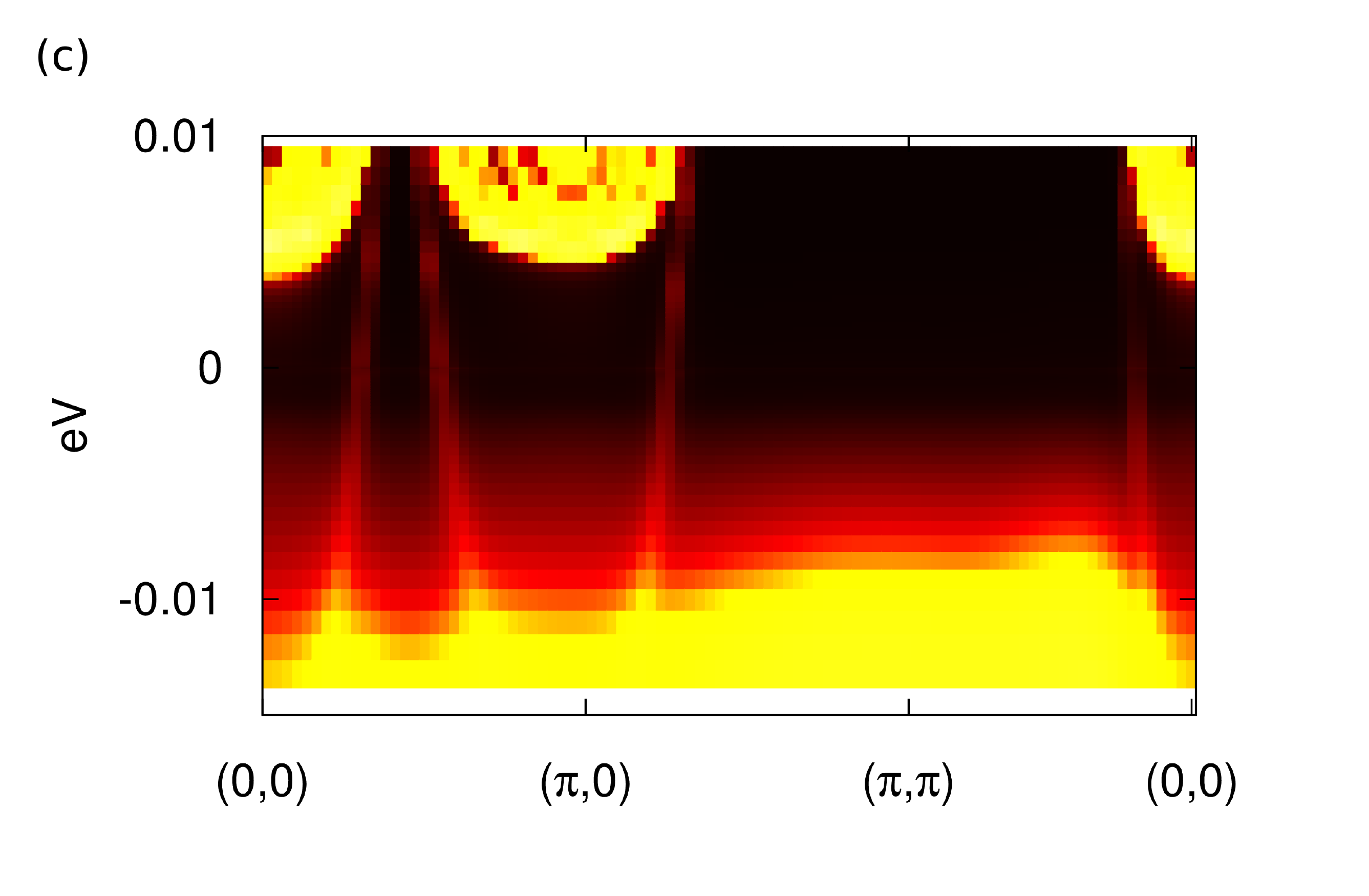}
\includegraphics[width=0.49\linewidth]{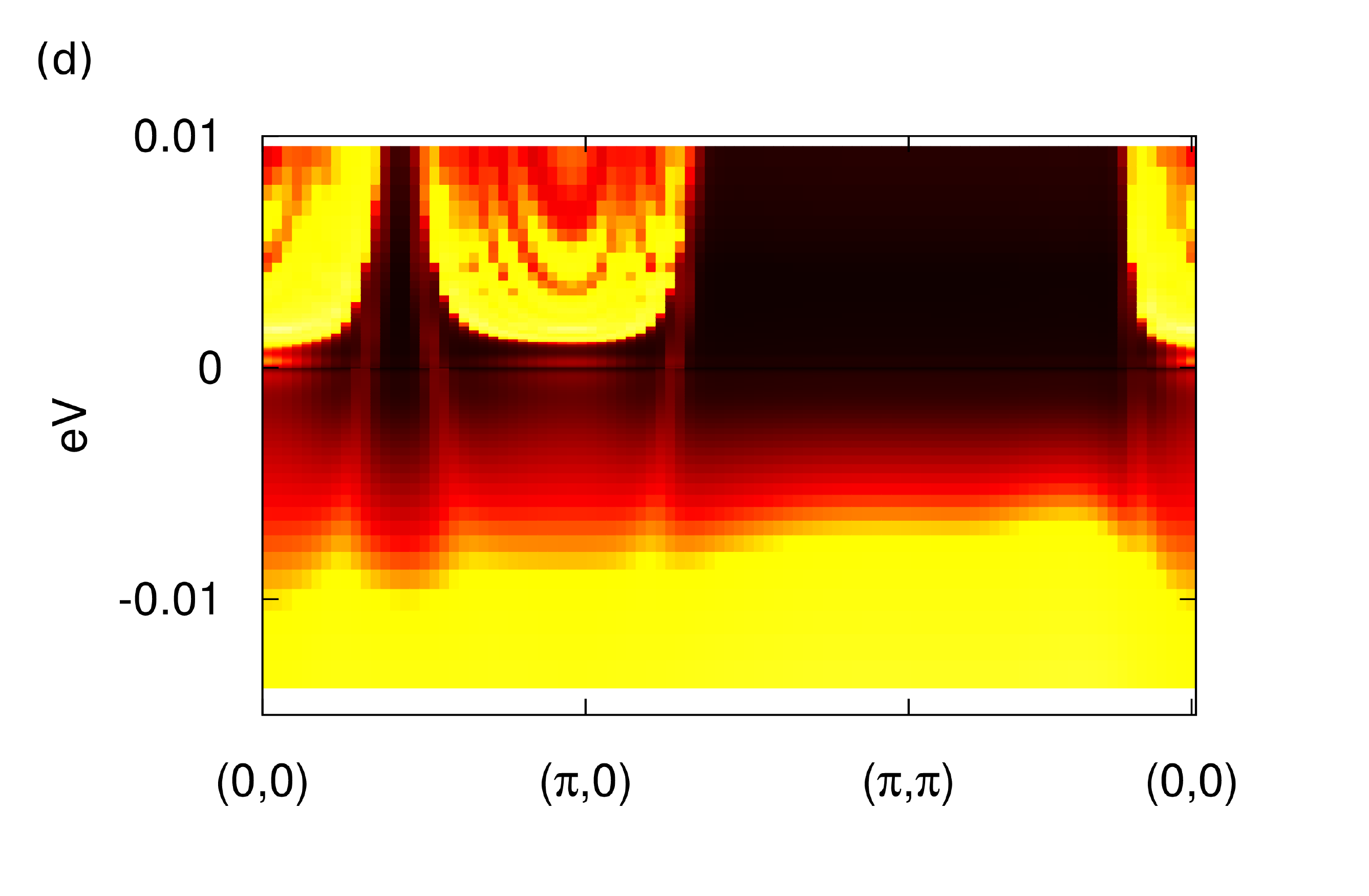}
\caption{(Color online) Momentum resolved spectral functions around the
  Fermi energy for different temperatures. Panel (a) $T=1K$, (b) $T=3K$, (c) $T=10K$, (d) $T=30K$.
\label{in_gap_T}}
\end{figure*}
Next, we want to analyze the effects of temperature on this strongly correlated topological state. The coexistence of light and heavy surface states rests in the formation of strongly layer-dependent Fermi liquids. If the temperature is increased above the coherence temperature of the outermost surface layer, the $f$-electrons in this layer will become incoherent. In Fig. \ref{in_gap_T} we show the combined momentum dependent spectrum of all layers for different temperatures. At $T=1K$, panel (a), the temperature is below the coherence temperature of the surface and the bulk, and we thus observe light and heavy surface states as described above. At $T=3K$, panel (b), the surface $f$-electrons become incoherent. At this temperature, the flat $f$-electron band, which is visible in panel (a), has vanished. Instead, only light surface states, which cross the Fermi energy, exist in the spectrum. This situation has been predicted in Alexandrov {\it et al.} \cite{PhysRevLett.114.177202}. When further increasing the temperature, the bulk gap begins slowly to close, see panel (c) and (d). For $T=30K$, panel (d), the width of the bulk gap has decreased below $\Delta=0.005$eV. However, the light surface states are still clearly visible at this temperature. For temperatures of approximately $T=50$K, the bulk gap is closed in our calculations and thus also the surface states vanish from the spectrum.

As observed by Alexandrov {\it et al.} \cite{PhysRevLett.114.177202}, the surface states exactly at the Fermi energy change from heavy to light, when the temperature is increased. However, as a further consequence of the large effective electron mass at the surface, the surface electrons become incoherent slightly away from the Fermi energy, which leads to the emergence of light surface states at low temperatures away from the Fermi energy. The imaginary part of the self-energy, which leads to the incoherence of the surface $f$-electrons away from the Fermi energy, is essential for the observation of this phenomenon.

\begin{figure}[t]
\begin{center}
\includegraphics[width=\linewidth]{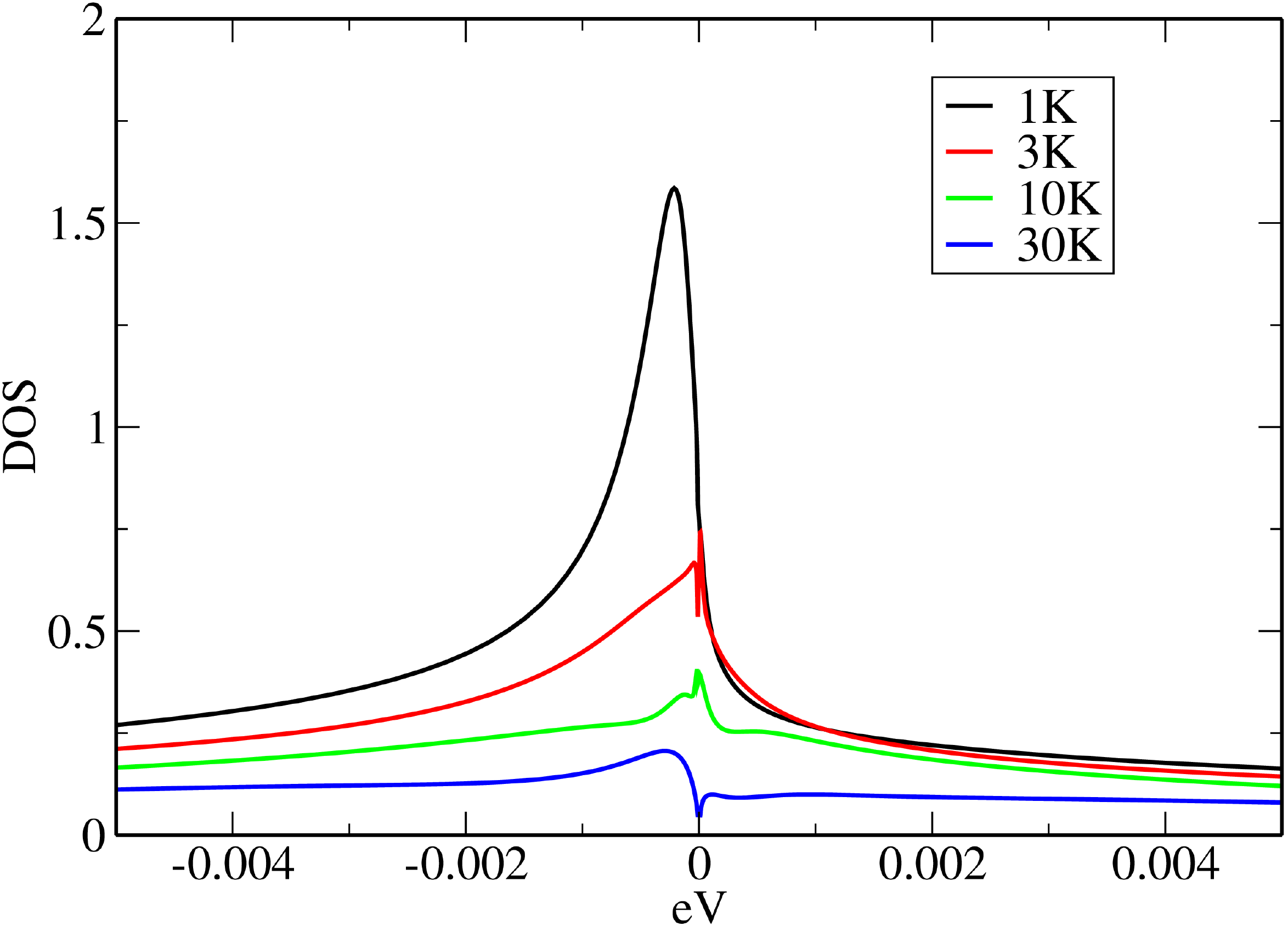}
\end{center}
\caption{Local Density of states of the $f$-electrons in the surface layer around the
  Fermi energy depending on the temperature.
\label{DOS_surface}}
\end{figure}
\begin{figure}[t]
\begin{center}
\includegraphics[width=\linewidth]{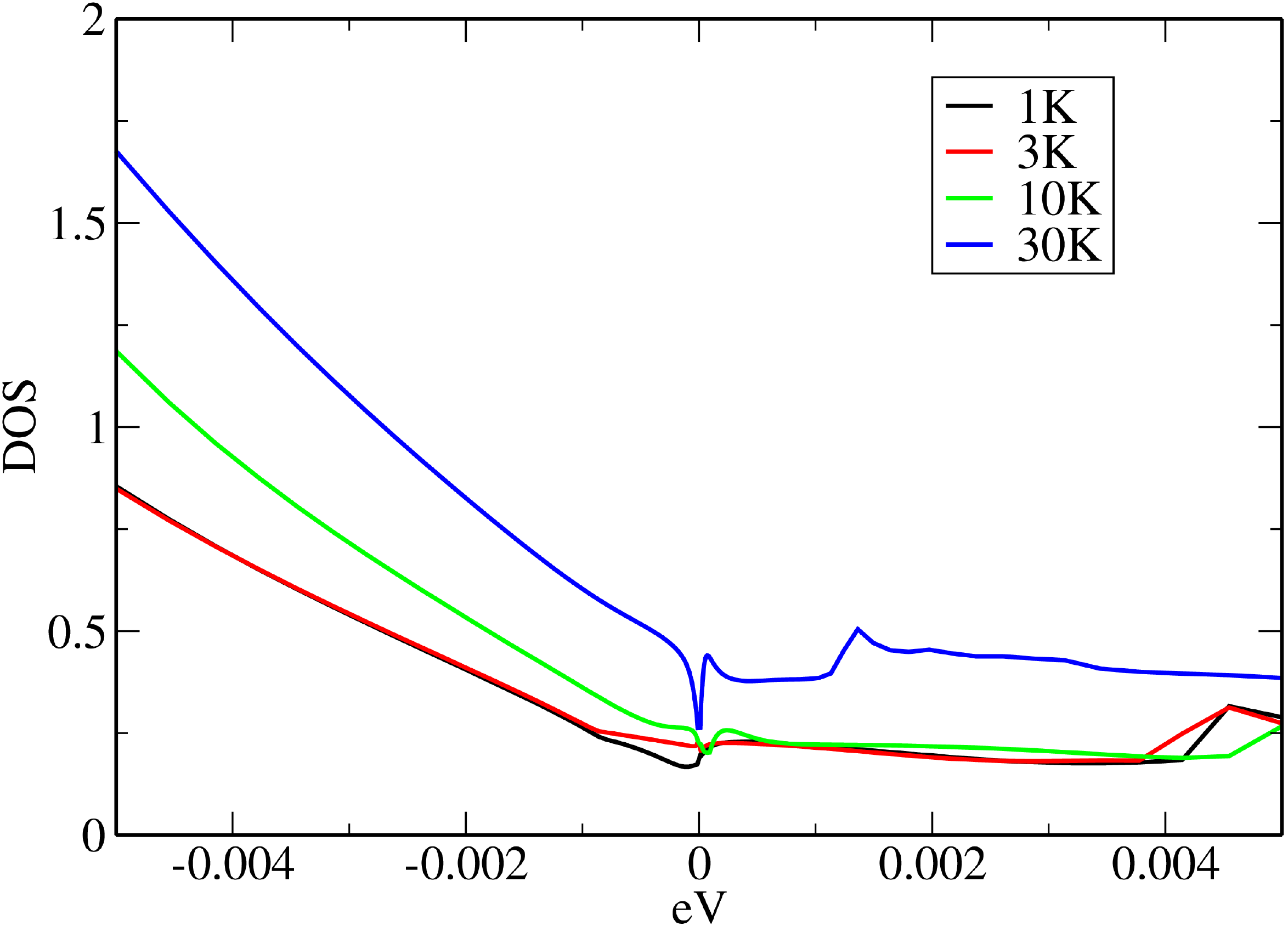}
\end{center}
\caption{Same as Fig. \ref{DOS_surface} but for the next-nearest surface layer.
\label{DOS_second}}
\end{figure}
The appearance of heavy surface states in the outermost layer and light surface states in the next-nearest surface layer can also be directly observed in the local density of states (DOS), Figs. \ref{DOS_surface} and \ref{DOS_second}. In the DOS of the surface layer, Fig. \ref{DOS_surface}, we observe at $T=1K$ a strong peak close to the Fermi energy, $\omega=0$. This peak corresponds to the flat band in the momentum-resolved spectrum close to the Fermi energy. The spectrum of the surface layer around the Fermi energy is strongly temperature dependent. Increasing the temperature, the $f$-electrons become incoherent and the peak in the DOS quickly vanishes, which leads to a decreasing spectral weight. However, the temperature dependence of the next-nearest surface layer shows the opposite behavior. When increasing the temperature, the DOS around the Fermi energy increases in this layer. This can be understood by two effects: (i) the $f$-electron band located at $\omega=-0.01$eV is broadened by the temperature, and (ii) due to the nontrivial topology, the light surface states in this layer exist in the whole gap.

\section{Conclusions}
We have analyzed the impact of strong correlations on the topological surface
states of a multi-band Kondo insulator.
We have elucidated a striking feature of the
combination of strong correlations and nontrivial topology: Due to a
strong increase of correlations at the surface, the $f$-electrons are
confined at the surface close to the Fermi energy forming heavy surface
states. However, because the bulk gap is larger than this region,
metallic states emerge in the next-nearest surface layer, which appear
as light surface states. Thus, the spectrum consists of  a combination
of light and heavy surface states. Furthermore we confirm
that when the temperature is
increased above the coherence temperature of the outermost layer, the
electrons of the surface layer become incoherent, so that the heavy
surface states vanish and the spectrum consists only of light surface states. This leads to an opposite temperature dependence of the local density of states of the surface layer and the next-nearest surface layer. While for increasing temperature, the DOS decreases in the surface layer around the Fermi energy, it increases in the next-nearest surface layer.

Finally we want to discuss the relevance of our results to SmB$_6$. First, our calculations have confirmed the previous results by Alexandrov {\it et al.} \cite{PhysRevLett.114.177202} about the Kondo breakdown scenario for SmB$_6$, which leads to a change from heavy surface states to light surface states, when the temperature is increased. Moreover, we have shown that the heavy topological surface states become light slightly away from the Fermi energy. This indicates that experiments need to access lower temperatures below the surface Kondo temperature and resolve small energies very close to the Fermi energy in order to observe heavy surface states. 

We note that our theoretical calculations in this paper have assumed a perfect surface of SmB$_6$, while some experiments have shown that the surface of SmB$_6$ is often reconstructed.\cite{Rossler2013,Yee2013} The surface of SmB$_6$ that is most frequently observed in experiments has a $1\times 2$ reconstructed structure, where one half of the Sm atoms are missing. This reconstruction will affect the band structure at the surface, and thus will modify the surface Kondo temperature. In Ref. \cite{Baruselli2015}, it has been pointed out that the reconstruction of the surface could decrease the surface Kondo temperature. This would have two effects: the characteristic temperature, below which heavy surface states emerge, would be decreased, and also the energy region around the Fermi energy, where the heavy surface states can be observed, would be reduced. Thus, due to surface reconstruction, it may be necessary to cool the samples down to lower temperatures, in order to experimentally observe the described interplay between non-trivial topology and strong correlations, which leads to a coexistence of heavy and light surface states.

In this paper, we have studied only Sm-terminated surfaces. Boron-terminated surfaces, which have also been analyzed in experiments and theory, will exhibit different surface properties. This issue will be addressed in a future study.

\begin{acknowledgments}
We appreciate helpful discussions with P. Coleman and X. Dai.
RP thanks for the support through the FPR program of
RIKEN. NK is supported through KAKENHI Grant No. 15H05855, 16K05501. Computer calculations have been done at the
supercomputer at RIKEN and the supercomputer of the ISSP in Japan.
\end{acknowledgments}

\bibliography{paper}

\begin{thebibliography}{61}
\expandafter\ifx\csname natexlab\endcsname\relax\def\natexlab#1{#1}\fi
\expandafter\ifx\csname bibnamefont\endcsname\relax
  \def\bibnamefont#1{#1}\fi
\expandafter\ifx\csname bibfnamefont\endcsname\relax
  \def\bibfnamefont#1{#1}\fi
\expandafter\ifx\csname citenamefont\endcsname\relax
  \def\citenamefont#1{#1}\fi
\expandafter\ifx\csname url\endcsname\relax
  \def\url#1{\texttt{#1}}\fi
\expandafter\ifx\csname urlprefix\endcsname\relax\def\urlprefix{URL }\fi
\providecommand{\bibinfo}[2]{#2}
\providecommand{\eprint}[2][]{\url{#2}}

\bibitem[{\citenamefont{Hasan and Kane}(2010)}]{RevModPhys.82.3045}
\bibinfo{author}{\bibfnamefont{M.~Z.} \bibnamefont{Hasan}} \bibnamefont{and}
  \bibinfo{author}{\bibfnamefont{C.~L.} \bibnamefont{Kane}},
  \bibinfo{journal}{Rev. Mod. Phys.} \textbf{\bibinfo{volume}{82}},
  \bibinfo{pages}{3045} (\bibinfo{year}{2010}),
  \urlprefix\url{http://link.aps.org/doi/10.1103/RevModPhys.82.3045}.

\bibitem[{\citenamefont{Qi and Zhang}(2011)}]{RevModPhys.83.1057}
\bibinfo{author}{\bibfnamefont{X.-L.} \bibnamefont{Qi}} \bibnamefont{and}
  \bibinfo{author}{\bibfnamefont{S.-C.} \bibnamefont{Zhang}},
  \bibinfo{journal}{Rev. Mod. Phys.} \textbf{\bibinfo{volume}{83}},
  \bibinfo{pages}{1057} (\bibinfo{year}{2011}),
  \urlprefix\url{http://link.aps.org/doi/10.1103/RevModPhys.83.1057}.

\bibitem[{\citenamefont{Pesin and Balents}(2010)}]{Pesin2010}
\bibinfo{author}{\bibfnamefont{D.}~\bibnamefont{Pesin}} \bibnamefont{and}
  \bibinfo{author}{\bibfnamefont{L.}~\bibnamefont{Balents}},
  \bibinfo{journal}{Nat Phys} \textbf{\bibinfo{volume}{6}},
  \bibinfo{pages}{376} (\bibinfo{year}{2010}), ISSN \bibinfo{issn}{1745-2473},
  \urlprefix\url{http://dx.doi.org/10.1038/nphys1606}.

\bibitem[{\citenamefont{Hohenadler et~al.}(2011)\citenamefont{Hohenadler, Lang,
  and Assaad}}]{PhysRevLett.106.100403}
\bibinfo{author}{\bibfnamefont{M.}~\bibnamefont{Hohenadler}},
  \bibinfo{author}{\bibfnamefont{T.~C.} \bibnamefont{Lang}}, \bibnamefont{and}
  \bibinfo{author}{\bibfnamefont{F.~F.} \bibnamefont{Assaad}},
  \bibinfo{journal}{Phys. Rev. Lett.} \textbf{\bibinfo{volume}{106}},
  \bibinfo{pages}{100403} (\bibinfo{year}{2011}),
  \urlprefix\url{http://link.aps.org/doi/10.1103/PhysRevLett.106.100403}.

\bibitem[{\citenamefont{Yu et~al.}(2011)\citenamefont{Yu, Xie, and
  Li}}]{PhysRevLett.107.010401}
\bibinfo{author}{\bibfnamefont{S.-L.} \bibnamefont{Yu}},
  \bibinfo{author}{\bibfnamefont{X.~C.} \bibnamefont{Xie}}, \bibnamefont{and}
  \bibinfo{author}{\bibfnamefont{J.-X.} \bibnamefont{Li}},
  \bibinfo{journal}{Phys. Rev. Lett.} \textbf{\bibinfo{volume}{107}},
  \bibinfo{pages}{010401} (\bibinfo{year}{2011}),
  \urlprefix\url{http://link.aps.org/doi/10.1103/PhysRevLett.107.010401}.

\bibitem[{\citenamefont{Hohenadler and Assaad}(2013)}]{0953-8984-25-14-143201}
\bibinfo{author}{\bibfnamefont{M.}~\bibnamefont{Hohenadler}} \bibnamefont{and}
  \bibinfo{author}{\bibfnamefont{F.~F.} \bibnamefont{Assaad}},
  \bibinfo{journal}{Journal of Physics: Condensed Matter}
  \textbf{\bibinfo{volume}{25}}, \bibinfo{pages}{143201}
  (\bibinfo{year}{2013}),
  \urlprefix\url{http://stacks.iop.org/0953-8984/25/i=14/a=143201}.

\bibitem[{\citenamefont{Yoshida et~al.}(2012)\citenamefont{Yoshida, Fujimoto,
  and Kawakami}}]{PhysRevB.85.125113}
\bibinfo{author}{\bibfnamefont{T.}~\bibnamefont{Yoshida}},
  \bibinfo{author}{\bibfnamefont{S.}~\bibnamefont{Fujimoto}}, \bibnamefont{and}
  \bibinfo{author}{\bibfnamefont{N.}~\bibnamefont{Kawakami}},
  \bibinfo{journal}{Phys. Rev. B} \textbf{\bibinfo{volume}{85}},
  \bibinfo{pages}{125113} (\bibinfo{year}{2012}),
  \urlprefix\url{http://link.aps.org/doi/10.1103/PhysRevB.85.125113}.

\bibitem[{\citenamefont{Fisk et~al.}(1995)\citenamefont{Fisk, Sarrao, Thompson,
  Mandrus, Hundley, Miglori, Bucher, Schlesinger, Aeppli, Bucher
  et~al.}}]{Fisk1995798}
\bibinfo{author}{\bibfnamefont{Z.}~\bibnamefont{Fisk}},
  \bibinfo{author}{\bibfnamefont{J.}~\bibnamefont{Sarrao}},
  \bibinfo{author}{\bibfnamefont{J.}~\bibnamefont{Thompson}},
  \bibinfo{author}{\bibfnamefont{D.}~\bibnamefont{Mandrus}},
  \bibinfo{author}{\bibfnamefont{M.}~\bibnamefont{Hundley}},
  \bibinfo{author}{\bibfnamefont{A.}~\bibnamefont{Miglori}},
  \bibinfo{author}{\bibfnamefont{B.}~\bibnamefont{Bucher}},
  \bibinfo{author}{\bibfnamefont{Z.}~\bibnamefont{Schlesinger}},
  \bibinfo{author}{\bibfnamefont{G.}~\bibnamefont{Aeppli}},
  \bibinfo{author}{\bibfnamefont{E.}~\bibnamefont{Bucher}},
  \bibnamefont{et~al.}, \bibinfo{journal}{Physica B: Condensed Matter}
  \textbf{\bibinfo{volume}{206–207}}, \bibinfo{pages}{798 }
  (\bibinfo{year}{1995}), ISSN \bibinfo{issn}{0921-4526},
  \bibinfo{note}{proceedings of the International Conference on Strongly
  Correlated Electron Systems},
  \urlprefix\url{http://www.sciencedirect.com/science/article/pii/092145269400588M}.

\bibitem[{\citenamefont{Coleman}(2007)}]{coleman2007}
\bibinfo{author}{\bibfnamefont{P.}~\bibnamefont{Coleman}},
  \emph{\bibinfo{title}{Handbook of Magnetism and Advanced Magnetic Materials}}
  (\bibinfo{publisher}{John Wiley and Sons}, \bibinfo{year}{2007}),
  p.~\bibinfo{pages}{95}.

\bibitem[{\citenamefont{Takimoto}(2011)}]{Takimoto2011}
\bibinfo{author}{\bibfnamefont{T.}~\bibnamefont{Takimoto}},
  \bibinfo{journal}{Journal of the Physical Society of Japan}
  \textbf{\bibinfo{volume}{80}}, \bibinfo{pages}{123710}
  (\bibinfo{year}{2011}), \eprint{http://dx.doi.org/10.1143/JPSJ.80.123710},
  \urlprefix\url{http://dx.doi.org/10.1143/JPSJ.80.123710}.

\bibitem[{\citenamefont{Yan et~al.}(2012)\citenamefont{Yan, M\"uchler, Qi,
  Zhang, and Felser}}]{PhysRevB.85.165125}
\bibinfo{author}{\bibfnamefont{B.}~\bibnamefont{Yan}},
  \bibinfo{author}{\bibfnamefont{L.}~\bibnamefont{M\"uchler}},
  \bibinfo{author}{\bibfnamefont{X.-L.} \bibnamefont{Qi}},
  \bibinfo{author}{\bibfnamefont{S.-C.} \bibnamefont{Zhang}}, \bibnamefont{and}
  \bibinfo{author}{\bibfnamefont{C.}~\bibnamefont{Felser}},
  \bibinfo{journal}{Phys. Rev. B} \textbf{\bibinfo{volume}{85}},
  \bibinfo{pages}{165125} (\bibinfo{year}{2012}),
  \urlprefix\url{http://link.aps.org/doi/10.1103/PhysRevB.85.165125}.

\bibitem[{\citenamefont{Menth et~al.}(1969)\citenamefont{Menth, Buehler, and
  Geballe}}]{PhysRevLett.22.295}
\bibinfo{author}{\bibfnamefont{A.}~\bibnamefont{Menth}},
  \bibinfo{author}{\bibfnamefont{E.}~\bibnamefont{Buehler}}, \bibnamefont{and}
  \bibinfo{author}{\bibfnamefont{T.~H.} \bibnamefont{Geballe}},
  \bibinfo{journal}{Phys. Rev. Lett.} \textbf{\bibinfo{volume}{22}},
  \bibinfo{pages}{295} (\bibinfo{year}{1969}),
  \urlprefix\url{http://link.aps.org/doi/10.1103/PhysRevLett.22.295}.

\bibitem[{\citenamefont{Allen et~al.}(1979)\citenamefont{Allen, Batlogg, and
  Wachter}}]{PhysRevB.20.4807}
\bibinfo{author}{\bibfnamefont{J.~W.} \bibnamefont{Allen}},
  \bibinfo{author}{\bibfnamefont{B.}~\bibnamefont{Batlogg}}, \bibnamefont{and}
  \bibinfo{author}{\bibfnamefont{P.}~\bibnamefont{Wachter}},
  \bibinfo{journal}{Phys. Rev. B} \textbf{\bibinfo{volume}{20}},
  \bibinfo{pages}{4807} (\bibinfo{year}{1979}),
  \urlprefix\url{http://link.aps.org/doi/10.1103/PhysRevB.20.4807}.

\bibitem[{\citenamefont{Wolgast et~al.}(2013)\citenamefont{Wolgast, Kurdak,
  Sun, Allen, Kim, and Fisk}}]{PhysRevB.88.180405}
\bibinfo{author}{\bibfnamefont{S.}~\bibnamefont{Wolgast}},
  \bibinfo{author}{\bibfnamefont{i.~m. c. b. u. i. e. i.~f.}
  \bibnamefont{Kurdak}}, \bibinfo{author}{\bibfnamefont{K.}~\bibnamefont{Sun}},
  \bibinfo{author}{\bibfnamefont{J.~W.} \bibnamefont{Allen}},
  \bibinfo{author}{\bibfnamefont{D.-J.} \bibnamefont{Kim}}, \bibnamefont{and}
  \bibinfo{author}{\bibfnamefont{Z.}~\bibnamefont{Fisk}},
  \bibinfo{journal}{Phys. Rev. B} \textbf{\bibinfo{volume}{88}},
  \bibinfo{pages}{180405} (\bibinfo{year}{2013}),
  \urlprefix\url{http://link.aps.org/doi/10.1103/PhysRevB.88.180405}.

\bibitem[{\citenamefont{Zhang et~al.}(2013)\citenamefont{Zhang, Butch, Syers,
  Ziemak, Greene, and Paglione}}]{PhysRevX.3.011011}
\bibinfo{author}{\bibfnamefont{X.}~\bibnamefont{Zhang}},
  \bibinfo{author}{\bibfnamefont{N.~P.} \bibnamefont{Butch}},
  \bibinfo{author}{\bibfnamefont{P.}~\bibnamefont{Syers}},
  \bibinfo{author}{\bibfnamefont{S.}~\bibnamefont{Ziemak}},
  \bibinfo{author}{\bibfnamefont{R.~L.} \bibnamefont{Greene}},
  \bibnamefont{and} \bibinfo{author}{\bibfnamefont{J.}~\bibnamefont{Paglione}},
  \bibinfo{journal}{Phys. Rev. X} \textbf{\bibinfo{volume}{3}},
  \bibinfo{pages}{011011} (\bibinfo{year}{2013}),
  \urlprefix\url{http://link.aps.org/doi/10.1103/PhysRevX.3.011011}.

\bibitem[{\citenamefont{Kim et~al.}(2013)\citenamefont{Kim, Thomas, Grant,
  Botimer, Fisk, and Xia}}]{Kim2013}
\bibinfo{author}{\bibfnamefont{D.~J.} \bibnamefont{Kim}},
  \bibinfo{author}{\bibfnamefont{S.}~\bibnamefont{Thomas}},
  \bibinfo{author}{\bibfnamefont{T.}~\bibnamefont{Grant}},
  \bibinfo{author}{\bibfnamefont{J.}~\bibnamefont{Botimer}},
  \bibinfo{author}{\bibfnamefont{Z.}~\bibnamefont{Fisk}}, \bibnamefont{and}
  \bibinfo{author}{\bibfnamefont{J.}~\bibnamefont{Xia}},
  \bibinfo{journal}{Scientific Reports} \textbf{\bibinfo{volume}{3}},
  \bibinfo{pages}{3150 EP } (\bibinfo{year}{2013}), \bibinfo{note}{article},
  \urlprefix\url{http://dx.doi.org/10.1038/srep03150}.

\bibitem[{\citenamefont{Li et~al.}(2014)\citenamefont{Li, Xiang, Yu, Asaba,
  Lawson, Cai, Tinsman, Berkley, Wolgast, Eo et~al.}}]{Li05122014}
\bibinfo{author}{\bibfnamefont{G.}~\bibnamefont{Li}},
  \bibinfo{author}{\bibfnamefont{Z.}~\bibnamefont{Xiang}},
  \bibinfo{author}{\bibfnamefont{F.}~\bibnamefont{Yu}},
  \bibinfo{author}{\bibfnamefont{T.}~\bibnamefont{Asaba}},
  \bibinfo{author}{\bibfnamefont{B.}~\bibnamefont{Lawson}},
  \bibinfo{author}{\bibfnamefont{P.}~\bibnamefont{Cai}},
  \bibinfo{author}{\bibfnamefont{C.}~\bibnamefont{Tinsman}},
  \bibinfo{author}{\bibfnamefont{A.}~\bibnamefont{Berkley}},
  \bibinfo{author}{\bibfnamefont{S.}~\bibnamefont{Wolgast}},
  \bibinfo{author}{\bibfnamefont{Y.~S.} \bibnamefont{Eo}},
  \bibnamefont{et~al.}, \bibinfo{journal}{Science}
  \textbf{\bibinfo{volume}{346}}, \bibinfo{pages}{1208} (\bibinfo{year}{2014}),
  \eprint{http://www.sciencemag.org/content/346/6214/1208.full.pdf},
  \urlprefix\url{http://www.sciencemag.org/content/346/6214/1208.abstract}.

\bibitem[{\citenamefont{Kim et~al.}(2014)\citenamefont{Kim, Xia, and
  Fisk}}]{Kim2014}
\bibinfo{author}{\bibfnamefont{D.~J.} \bibnamefont{Kim}},
  \bibinfo{author}{\bibfnamefont{J.}~\bibnamefont{Xia}}, \bibnamefont{and}
  \bibinfo{author}{\bibfnamefont{Z.}~\bibnamefont{Fisk}}, \bibinfo{journal}{Nat
  Mater} \textbf{\bibinfo{volume}{13}}, \bibinfo{pages}{466}
  (\bibinfo{year}{2014}), ISSN \bibinfo{issn}{1476-1122},
  \bibinfo{note}{letter}, \urlprefix\url{http://dx.doi.org/10.1038/nmat3913}.

\bibitem[{\citenamefont{Dzero et~al.}(2010)\citenamefont{Dzero, Sun, Galitski,
  and Coleman}}]{PhysRevLett.104.106408}
\bibinfo{author}{\bibfnamefont{M.}~\bibnamefont{Dzero}},
  \bibinfo{author}{\bibfnamefont{K.}~\bibnamefont{Sun}},
  \bibinfo{author}{\bibfnamefont{V.}~\bibnamefont{Galitski}}, \bibnamefont{and}
  \bibinfo{author}{\bibfnamefont{P.}~\bibnamefont{Coleman}},
  \bibinfo{journal}{Phys. Rev. Lett.} \textbf{\bibinfo{volume}{104}},
  \bibinfo{pages}{106408} (\bibinfo{year}{2010}),
  \urlprefix\url{http://link.aps.org/doi/10.1103/PhysRevLett.104.106408}.

\bibitem[{\citenamefont{Dzero et~al.}(2012)\citenamefont{Dzero, Sun, Coleman,
  and Galitski}}]{PhysRevB.85.045130}
\bibinfo{author}{\bibfnamefont{M.}~\bibnamefont{Dzero}},
  \bibinfo{author}{\bibfnamefont{K.}~\bibnamefont{Sun}},
  \bibinfo{author}{\bibfnamefont{P.}~\bibnamefont{Coleman}}, \bibnamefont{and}
  \bibinfo{author}{\bibfnamefont{V.}~\bibnamefont{Galitski}},
  \bibinfo{journal}{Phys. Rev. B} \textbf{\bibinfo{volume}{85}},
  \bibinfo{pages}{045130} (\bibinfo{year}{2012}),
  \urlprefix\url{http://link.aps.org/doi/10.1103/PhysRevB.85.045130}.

\bibitem[{\citenamefont{Alexandrov et~al.}(2013)\citenamefont{Alexandrov,
  Dzero, and Coleman}}]{PhysRevLett.111.226403}
\bibinfo{author}{\bibfnamefont{V.}~\bibnamefont{Alexandrov}},
  \bibinfo{author}{\bibfnamefont{M.}~\bibnamefont{Dzero}}, \bibnamefont{and}
  \bibinfo{author}{\bibfnamefont{P.}~\bibnamefont{Coleman}},
  \bibinfo{journal}{Phys. Rev. Lett.} \textbf{\bibinfo{volume}{111}},
  \bibinfo{pages}{226403} (\bibinfo{year}{2013}),
  \urlprefix\url{http://link.aps.org/doi/10.1103/PhysRevLett.111.226403}.

\bibitem[{\citenamefont{Lu et~al.}(2013)\citenamefont{Lu, Zhao, Weng, Fang, and
  Dai}}]{PhysRevLett.110.096401}
\bibinfo{author}{\bibfnamefont{F.}~\bibnamefont{Lu}},
  \bibinfo{author}{\bibfnamefont{J.}~\bibnamefont{Zhao}},
  \bibinfo{author}{\bibfnamefont{H.}~\bibnamefont{Weng}},
  \bibinfo{author}{\bibfnamefont{Z.}~\bibnamefont{Fang}}, \bibnamefont{and}
  \bibinfo{author}{\bibfnamefont{X.}~\bibnamefont{Dai}},
  \bibinfo{journal}{Phys. Rev. Lett.} \textbf{\bibinfo{volume}{110}},
  \bibinfo{pages}{096401} (\bibinfo{year}{2013}),
  \urlprefix\url{http://link.aps.org/doi/10.1103/PhysRevLett.110.096401}.

\bibitem[{\citenamefont{Tran et~al.}(2012)\citenamefont{Tran, Takimoto, and
  Kim}}]{PhysRevB.85.125128}
\bibinfo{author}{\bibfnamefont{M.-T.} \bibnamefont{Tran}},
  \bibinfo{author}{\bibfnamefont{T.}~\bibnamefont{Takimoto}}, \bibnamefont{and}
  \bibinfo{author}{\bibfnamefont{K.-S.} \bibnamefont{Kim}},
  \bibinfo{journal}{Phys. Rev. B} \textbf{\bibinfo{volume}{85}},
  \bibinfo{pages}{125128} (\bibinfo{year}{2012}),
  \urlprefix\url{http://link.aps.org/doi/10.1103/PhysRevB.85.125128}.

\bibitem[{\citenamefont{Werner and Assaad}(2013)}]{PhysRevB.88.035113}
\bibinfo{author}{\bibfnamefont{J.}~\bibnamefont{Werner}} \bibnamefont{and}
  \bibinfo{author}{\bibfnamefont{F.~F.} \bibnamefont{Assaad}},
  \bibinfo{journal}{Phys. Rev. B} \textbf{\bibinfo{volume}{88}},
  \bibinfo{pages}{035113} (\bibinfo{year}{2013}),
  \urlprefix\url{http://link.aps.org/doi/10.1103/PhysRevB.88.035113}.

\bibitem[{\citenamefont{Legner et~al.}(2014)\citenamefont{Legner, R\"uegg, and
  Sigrist}}]{PhysRevB.89.085110}
\bibinfo{author}{\bibfnamefont{M.}~\bibnamefont{Legner}},
  \bibinfo{author}{\bibfnamefont{A.}~\bibnamefont{R\"uegg}}, \bibnamefont{and}
  \bibinfo{author}{\bibfnamefont{M.}~\bibnamefont{Sigrist}},
  \bibinfo{journal}{Phys. Rev. B} \textbf{\bibinfo{volume}{89}},
  \bibinfo{pages}{085110} (\bibinfo{year}{2014}),
  \urlprefix\url{http://link.aps.org/doi/10.1103/PhysRevB.89.085110}.

\bibitem[{\citenamefont{Werner and Assaad}(2014)}]{PhysRevB.89.245119}
\bibinfo{author}{\bibfnamefont{J.}~\bibnamefont{Werner}} \bibnamefont{and}
  \bibinfo{author}{\bibfnamefont{F.~F.} \bibnamefont{Assaad}},
  \bibinfo{journal}{Phys. Rev. B} \textbf{\bibinfo{volume}{89}},
  \bibinfo{pages}{245119} (\bibinfo{year}{2014}),
  \urlprefix\url{http://link.aps.org/doi/10.1103/PhysRevB.89.245119}.

\bibitem[{\citenamefont{Efimkin and Galitski}(2014)}]{PhysRevB.90.081113}
\bibinfo{author}{\bibfnamefont{D.~K.} \bibnamefont{Efimkin}} \bibnamefont{and}
  \bibinfo{author}{\bibfnamefont{V.}~\bibnamefont{Galitski}},
  \bibinfo{journal}{Phys. Rev. B} \textbf{\bibinfo{volume}{90}},
  \bibinfo{pages}{081113} (\bibinfo{year}{2014}),
  \urlprefix\url{http://link.aps.org/doi/10.1103/PhysRevB.90.081113}.

\bibitem[{\citenamefont{Fuhrman and Nikoli\ifmmode~\acute{c}\else
  \'{c}\fi{}}(2014)}]{PhysRevB.90.195144}
\bibinfo{author}{\bibfnamefont{W.~T.} \bibnamefont{Fuhrman}} \bibnamefont{and}
  \bibinfo{author}{\bibfnamefont{P.}~\bibnamefont{Nikoli\ifmmode~\acute{c}\else
  \'{c}\fi{}}}, \bibinfo{journal}{Phys. Rev. B} \textbf{\bibinfo{volume}{90}},
  \bibinfo{pages}{195144} (\bibinfo{year}{2014}),
  \urlprefix\url{http://link.aps.org/doi/10.1103/PhysRevB.90.195144}.

\bibitem[{\citenamefont{Nikoli\ifmmode~\acute{c}\else
  \'{c}\fi{}}(2014)}]{PhysRevB.90.235107}
\bibinfo{author}{\bibfnamefont{P.}~\bibnamefont{Nikoli\ifmmode~\acute{c}\else
  \'{c}\fi{}}}, \bibinfo{journal}{Phys. Rev. B} \textbf{\bibinfo{volume}{90}},
  \bibinfo{pages}{235107} (\bibinfo{year}{2014}),
  \urlprefix\url{http://link.aps.org/doi/10.1103/PhysRevB.90.235107}.

\bibitem[{\citenamefont{Schlottmann}(2014)}]{PhysRevB.90.165127}
\bibinfo{author}{\bibfnamefont{P.}~\bibnamefont{Schlottmann}},
  \bibinfo{journal}{Phys. Rev. B} \textbf{\bibinfo{volume}{90}},
  \bibinfo{pages}{165127} (\bibinfo{year}{2014}),
  \urlprefix\url{http://link.aps.org/doi/10.1103/PhysRevB.90.165127}.

\bibitem[{\citenamefont{Roy et~al.}(2014)\citenamefont{Roy, Sau, Dzero, and
  Galitski}}]{PhysRevB.90.155314}
\bibinfo{author}{\bibfnamefont{B.}~\bibnamefont{Roy}},
  \bibinfo{author}{\bibfnamefont{J.~D.} \bibnamefont{Sau}},
  \bibinfo{author}{\bibfnamefont{M.}~\bibnamefont{Dzero}}, \bibnamefont{and}
  \bibinfo{author}{\bibfnamefont{V.}~\bibnamefont{Galitski}},
  \bibinfo{journal}{Phys. Rev. B} \textbf{\bibinfo{volume}{90}},
  \bibinfo{pages}{155314} (\bibinfo{year}{2014}),
  \urlprefix\url{http://link.aps.org/doi/10.1103/PhysRevB.90.155314}.

\bibitem[{\citenamefont{Chen et~al.}(2014)\citenamefont{Chen, Werner, and
  Assaad}}]{PhysRevB.90.115109}
\bibinfo{author}{\bibfnamefont{K.-S.} \bibnamefont{Chen}},
  \bibinfo{author}{\bibfnamefont{J.}~\bibnamefont{Werner}}, \bibnamefont{and}
  \bibinfo{author}{\bibfnamefont{F.}~\bibnamefont{Assaad}},
  \bibinfo{journal}{Phys. Rev. B} \textbf{\bibinfo{volume}{90}},
  \bibinfo{pages}{115109} (\bibinfo{year}{2014}),
  \urlprefix\url{http://link.aps.org/doi/10.1103/PhysRevB.90.115109}.

\bibitem[{\citenamefont{Baruselli and Vojta}(2014)}]{PhysRevB.90.201106}
\bibinfo{author}{\bibfnamefont{P.~P.} \bibnamefont{Baruselli}}
  \bibnamefont{and} \bibinfo{author}{\bibfnamefont{M.}~\bibnamefont{Vojta}},
  \bibinfo{journal}{Phys. Rev. B} \textbf{\bibinfo{volume}{90}},
  \bibinfo{pages}{201106} (\bibinfo{year}{2014}),
  \urlprefix\url{http://link.aps.org/doi/10.1103/PhysRevB.90.201106}.

\bibitem[{\citenamefont{Iaconis and Balents}(2015)}]{PhysRevB.91.245127}
\bibinfo{author}{\bibfnamefont{J.}~\bibnamefont{Iaconis}} \bibnamefont{and}
  \bibinfo{author}{\bibfnamefont{L.}~\bibnamefont{Balents}},
  \bibinfo{journal}{Phys. Rev. B} \textbf{\bibinfo{volume}{91}},
  \bibinfo{pages}{245127} (\bibinfo{year}{2015}),
  \urlprefix\url{http://link.aps.org/doi/10.1103/PhysRevB.91.245127}.

\bibitem[{\citenamefont{Min et~al.}(2014)\citenamefont{Min, Lutz, Fiedler,
  Kang, Cho, Kim, Bentmann, and Reinert}}]{PhysRevLett.112.226402}
\bibinfo{author}{\bibfnamefont{C.-H.} \bibnamefont{Min}},
  \bibinfo{author}{\bibfnamefont{P.}~\bibnamefont{Lutz}},
  \bibinfo{author}{\bibfnamefont{S.}~\bibnamefont{Fiedler}},
  \bibinfo{author}{\bibfnamefont{B.~Y.} \bibnamefont{Kang}},
  \bibinfo{author}{\bibfnamefont{B.~K.} \bibnamefont{Cho}},
  \bibinfo{author}{\bibfnamefont{H.-D.} \bibnamefont{Kim}},
  \bibinfo{author}{\bibfnamefont{H.}~\bibnamefont{Bentmann}}, \bibnamefont{and}
  \bibinfo{author}{\bibfnamefont{F.}~\bibnamefont{Reinert}},
  \bibinfo{journal}{Phys. Rev. Lett.} \textbf{\bibinfo{volume}{112}},
  \bibinfo{pages}{226402} (\bibinfo{year}{2014}),
  \urlprefix\url{http://link.aps.org/doi/10.1103/PhysRevLett.112.226402}.

\bibitem[{\citenamefont{Yoshida et~al.}(2013)\citenamefont{Yoshida, Peters,
  Fujimoto, and Kawakami}}]{PhysRevB.87.165109}
\bibinfo{author}{\bibfnamefont{T.}~\bibnamefont{Yoshida}},
  \bibinfo{author}{\bibfnamefont{R.}~\bibnamefont{Peters}},
  \bibinfo{author}{\bibfnamefont{S.}~\bibnamefont{Fujimoto}}, \bibnamefont{and}
  \bibinfo{author}{\bibfnamefont{N.}~\bibnamefont{Kawakami}},
  \bibinfo{journal}{Phys. Rev. B} \textbf{\bibinfo{volume}{87}},
  \bibinfo{pages}{165109} (\bibinfo{year}{2013}),
  \urlprefix\url{http://link.aps.org/doi/10.1103/PhysRevB.87.165109}.

\bibitem[{\citenamefont{Yoshida et~al.}(2015)\citenamefont{Yoshida, Peters, and
  Kawakami}}]{yoshida2015}
\bibinfo{author}{\bibfnamefont{T.}~\bibnamefont{Yoshida}},
  \bibinfo{author}{\bibfnamefont{R.}~\bibnamefont{Peters}}, \bibnamefont{and}
  \bibinfo{author}{\bibfnamefont{N.}~\bibnamefont{Kawakami}}
  (\bibinfo{year}{2015}), \eprint{arXiv:1508.07779}.

\bibitem[{\citenamefont{Jiang et~al.}(2013)\citenamefont{Jiang, Li, Zhang, Sun,
  Chen, Ye, Xu, Ge, Tan, Niu et~al.}}]{Jiang2013}
\bibinfo{author}{\bibfnamefont{J.}~\bibnamefont{Jiang}},
  \bibinfo{author}{\bibfnamefont{S.}~\bibnamefont{Li}},
  \bibinfo{author}{\bibfnamefont{T.}~\bibnamefont{Zhang}},
  \bibinfo{author}{\bibfnamefont{Z.}~\bibnamefont{Sun}},
  \bibinfo{author}{\bibfnamefont{F.}~\bibnamefont{Chen}},
  \bibinfo{author}{\bibfnamefont{Z.~R.} \bibnamefont{Ye}},
  \bibinfo{author}{\bibfnamefont{M.}~\bibnamefont{Xu}},
  \bibinfo{author}{\bibfnamefont{Q.~Q.} \bibnamefont{Ge}},
  \bibinfo{author}{\bibfnamefont{S.~Y.} \bibnamefont{Tan}},
  \bibinfo{author}{\bibfnamefont{X.~H.} \bibnamefont{Niu}},
  \bibnamefont{et~al.}, \bibinfo{journal}{Nat Commun}
  \textbf{\bibinfo{volume}{4}} (\bibinfo{year}{2013}), \bibinfo{note}{article},
  \urlprefix\url{http://dx.doi.org/10.1038/ncomms4010}.

\bibitem[{\citenamefont{Neupane et~al.}(2013)\citenamefont{Neupane, Alidoust,
  Xu, Kondo, Ishida, Kim, Liu, Belopolski, Jo, Chang et~al.}}]{Neupane2013}
\bibinfo{author}{\bibfnamefont{M.}~\bibnamefont{Neupane}},
  \bibinfo{author}{\bibfnamefont{N.}~\bibnamefont{Alidoust}},
  \bibinfo{author}{\bibfnamefont{S.-Y.} \bibnamefont{Xu}},
  \bibinfo{author}{\bibfnamefont{T.}~\bibnamefont{Kondo}},
  \bibinfo{author}{\bibfnamefont{Y.}~\bibnamefont{Ishida}},
  \bibinfo{author}{\bibfnamefont{D.~J.} \bibnamefont{Kim}},
  \bibinfo{author}{\bibfnamefont{C.}~\bibnamefont{Liu}},
  \bibinfo{author}{\bibfnamefont{I.}~\bibnamefont{Belopolski}},
  \bibinfo{author}{\bibfnamefont{Y.~J.} \bibnamefont{Jo}},
  \bibinfo{author}{\bibfnamefont{T.-R.} \bibnamefont{Chang}},
  \bibnamefont{et~al.}, \bibinfo{journal}{Nat Commun}
  \textbf{\bibinfo{volume}{4}} (\bibinfo{year}{2013}),
  \urlprefix\url{http://dx.doi.org/10.1038/ncomms3991}.

\bibitem[{\citenamefont{Xu et~al.}(2013)\citenamefont{Xu, Shi, Biswas, Matt,
  Dhaka, Huang, Plumb, Radovi\ifmmode~\acute{c}\else \'{c}\fi{}, Dil,
  Pomjakushina et~al.}}]{PhysRevB.88.121102}
\bibinfo{author}{\bibfnamefont{N.}~\bibnamefont{Xu}},
  \bibinfo{author}{\bibfnamefont{X.}~\bibnamefont{Shi}},
  \bibinfo{author}{\bibfnamefont{P.~K.} \bibnamefont{Biswas}},
  \bibinfo{author}{\bibfnamefont{C.~E.} \bibnamefont{Matt}},
  \bibinfo{author}{\bibfnamefont{R.~S.} \bibnamefont{Dhaka}},
  \bibinfo{author}{\bibfnamefont{Y.}~\bibnamefont{Huang}},
  \bibinfo{author}{\bibfnamefont{N.~C.} \bibnamefont{Plumb}},
  \bibinfo{author}{\bibfnamefont{M.}~\bibnamefont{Radovi\ifmmode~\acute{c}\else
  \'{c}\fi{}}}, \bibinfo{author}{\bibfnamefont{J.~H.} \bibnamefont{Dil}},
  \bibinfo{author}{\bibfnamefont{E.}~\bibnamefont{Pomjakushina}},
  \bibnamefont{et~al.}, \bibinfo{journal}{Phys. Rev. B}
  \textbf{\bibinfo{volume}{88}}, \bibinfo{pages}{121102}
  (\bibinfo{year}{2013}),
  \urlprefix\url{http://link.aps.org/doi/10.1103/PhysRevB.88.121102}.

\bibitem[{\citenamefont{Zhu et~al.}(2013)\citenamefont{Zhu, Nicolaou, Levy,
  Butch, Syers, Wang, Paglione, Sawatzky, Elfimov, and
  Damascelli}}]{PhysRevLett.111.216402}
\bibinfo{author}{\bibfnamefont{Z.-H.} \bibnamefont{Zhu}},
  \bibinfo{author}{\bibfnamefont{A.}~\bibnamefont{Nicolaou}},
  \bibinfo{author}{\bibfnamefont{G.}~\bibnamefont{Levy}},
  \bibinfo{author}{\bibfnamefont{N.~P.} \bibnamefont{Butch}},
  \bibinfo{author}{\bibfnamefont{P.}~\bibnamefont{Syers}},
  \bibinfo{author}{\bibfnamefont{X.~F.} \bibnamefont{Wang}},
  \bibinfo{author}{\bibfnamefont{J.}~\bibnamefont{Paglione}},
  \bibinfo{author}{\bibfnamefont{G.~A.} \bibnamefont{Sawatzky}},
  \bibinfo{author}{\bibfnamefont{I.~S.} \bibnamefont{Elfimov}},
  \bibnamefont{and}
  \bibinfo{author}{\bibfnamefont{A.}~\bibnamefont{Damascelli}},
  \bibinfo{journal}{Phys. Rev. Lett.} \textbf{\bibinfo{volume}{111}},
  \bibinfo{pages}{216402} (\bibinfo{year}{2013}),
  \urlprefix\url{http://link.aps.org/doi/10.1103/PhysRevLett.111.216402}.

\bibitem[{\citenamefont{Frantzeskakis et~al.}(2013)\citenamefont{Frantzeskakis,
  de~Jong, Zwartsenberg, Huang, Pan, Zhang, Zhang, Zhang, Bao, Tegus
  et~al.}}]{PhysRevX.3.041024}
\bibinfo{author}{\bibfnamefont{E.}~\bibnamefont{Frantzeskakis}},
  \bibinfo{author}{\bibfnamefont{N.}~\bibnamefont{de~Jong}},
  \bibinfo{author}{\bibfnamefont{B.}~\bibnamefont{Zwartsenberg}},
  \bibinfo{author}{\bibfnamefont{Y.~K.} \bibnamefont{Huang}},
  \bibinfo{author}{\bibfnamefont{Y.}~\bibnamefont{Pan}},
  \bibinfo{author}{\bibfnamefont{X.}~\bibnamefont{Zhang}},
  \bibinfo{author}{\bibfnamefont{J.~X.} \bibnamefont{Zhang}},
  \bibinfo{author}{\bibfnamefont{F.~X.} \bibnamefont{Zhang}},
  \bibinfo{author}{\bibfnamefont{L.~H.} \bibnamefont{Bao}},
  \bibinfo{author}{\bibfnamefont{O.}~\bibnamefont{Tegus}},
  \bibnamefont{et~al.}, \bibinfo{journal}{Phys. Rev. X}
  \textbf{\bibinfo{volume}{3}}, \bibinfo{pages}{041024} (\bibinfo{year}{2013}),
  \urlprefix\url{http://link.aps.org/doi/10.1103/PhysRevX.3.041024}.

\bibitem[{\citenamefont{Xu et~al.}(2014)\citenamefont{Xu, Biswas, Dil, Dhaka,
  Landolt, Muff, Matt, Shi, Plumb, Radovi{\"A}? et~al.}}]{Xu2014}
\bibinfo{author}{\bibfnamefont{N.}~\bibnamefont{Xu}},
  \bibinfo{author}{\bibfnamefont{P.~K.} \bibnamefont{Biswas}},
  \bibinfo{author}{\bibfnamefont{J.~H.} \bibnamefont{Dil}},
  \bibinfo{author}{\bibfnamefont{R.~S.} \bibnamefont{Dhaka}},
  \bibinfo{author}{\bibfnamefont{G.}~\bibnamefont{Landolt}},
  \bibinfo{author}{\bibfnamefont{S.}~\bibnamefont{Muff}},
  \bibinfo{author}{\bibfnamefont{C.~E.} \bibnamefont{Matt}},
  \bibinfo{author}{\bibfnamefont{X.}~\bibnamefont{Shi}},
  \bibinfo{author}{\bibfnamefont{N.~C.} \bibnamefont{Plumb}},
  \bibinfo{author}{\bibfnamefont{M.}~\bibnamefont{Radovi{\"A}?}},
  \bibnamefont{et~al.}, \bibinfo{journal}{Nat Commun}
  \textbf{\bibinfo{volume}{5}} (\bibinfo{year}{2014}), \bibinfo{note}{article},
  \urlprefix\url{http://dx.doi.org/10.1038/ncomms5566}.

\bibitem[{\citenamefont{Alexandrov et~al.}(2015)\citenamefont{Alexandrov,
  Coleman, and Erten}}]{PhysRevLett.114.177202}
\bibinfo{author}{\bibfnamefont{V.}~\bibnamefont{Alexandrov}},
  \bibinfo{author}{\bibfnamefont{P.}~\bibnamefont{Coleman}}, \bibnamefont{and}
  \bibinfo{author}{\bibfnamefont{O.}~\bibnamefont{Erten}},
  \bibinfo{journal}{Phys. Rev. Lett.} \textbf{\bibinfo{volume}{114}},
  \bibinfo{pages}{177202} (\bibinfo{year}{2015}),
  \urlprefix\url{http://link.aps.org/doi/10.1103/PhysRevLett.114.177202}.

\bibitem[{\citenamefont{Georges et~al.}(1996)\citenamefont{Georges, Kotliar,
  Krauth, and Rozenberg}}]{Georges1996}
\bibinfo{author}{\bibfnamefont{A.}~\bibnamefont{Georges}},
  \bibinfo{author}{\bibfnamefont{G.}~\bibnamefont{Kotliar}},
  \bibinfo{author}{\bibfnamefont{W.}~\bibnamefont{Krauth}}, \bibnamefont{and}
  \bibinfo{author}{\bibfnamefont{M.~J.} \bibnamefont{Rozenberg}},
  \bibinfo{journal}{Rev. Mod. Phys.} \textbf{\bibinfo{volume}{68}},
  \bibinfo{pages}{13} (\bibinfo{year}{1996}),
  \urlprefix\url{http://link.aps.org/doi/10.1103/RevModPhys.68.13}.

\bibitem[{\citenamefont{Blaha et~al.}(2001)\citenamefont{Blaha, Schwarz,
  Madsen, Kvasnicka, and Luit}}]{Wien2K}
\bibinfo{author}{\bibfnamefont{P.}~\bibnamefont{Blaha}},
  \bibinfo{author}{\bibfnamefont{K.}~\bibnamefont{Schwarz}},
  \bibinfo{author}{\bibfnamefont{G.~K.~H.} \bibnamefont{Madsen}},
  \bibinfo{author}{\bibfnamefont{D.}~\bibnamefont{Kvasnicka}},
  \bibnamefont{and} \bibinfo{author}{\bibfnamefont{J.}~\bibnamefont{Luit}},
  \bibinfo{journal}{Wien2k: An Augmented Plane Wave + Local Orbitals Program
  for Calculating Crystal Properties (Vienna University of Technology, Wien)}
  (\bibinfo{year}{2001}), \bibinfo{note}{(We take $12\times 12\times 12$
  $k$-point meshes and the largest plane wave vector $K_{\rm max}$ as $R_{\rm
  MT} K_{\rm max}=7.0$, where $R_{\rm MT}$ is the muffin-tin radius(2.5 bohr
  for Sm and 1.55 bohr for B atoms).)}.

\bibitem[{\citenamefont{Kuneš et~al.}(2010)\citenamefont{Kuneš, Arita,
  Wissgott, Toschi, Ikeda, and Held}}]{Kunes20101888}
\bibinfo{author}{\bibfnamefont{J.}~\bibnamefont{Kuneš}},
  \bibinfo{author}{\bibfnamefont{R.}~\bibnamefont{Arita}},
  \bibinfo{author}{\bibfnamefont{P.}~\bibnamefont{Wissgott}},
  \bibinfo{author}{\bibfnamefont{A.}~\bibnamefont{Toschi}},
  \bibinfo{author}{\bibfnamefont{H.}~\bibnamefont{Ikeda}}, \bibnamefont{and}
  \bibinfo{author}{\bibfnamefont{K.}~\bibnamefont{Held}},
  \bibinfo{journal}{Computer Physics Communications}
  \textbf{\bibinfo{volume}{181}}, \bibinfo{pages}{1888 }
  (\bibinfo{year}{2010}), ISSN \bibinfo{issn}{0010-4655},
  \urlprefix\url{http://www.sciencedirect.com/science/article/pii/S0010465510002948}.

\bibitem[{\citenamefont{Marzari and Vanderbilt}(1997)}]{PhysRevB.56.12847}
\bibinfo{author}{\bibfnamefont{N.}~\bibnamefont{Marzari}} \bibnamefont{and}
  \bibinfo{author}{\bibfnamefont{D.}~\bibnamefont{Vanderbilt}},
  \bibinfo{journal}{Phys. Rev. B} \textbf{\bibinfo{volume}{56}},
  \bibinfo{pages}{12847} (\bibinfo{year}{1997}),
  \urlprefix\url{http://link.aps.org/doi/10.1103/PhysRevB.56.12847}.

\bibitem[{\citenamefont{Souza et~al.}(2001)\citenamefont{Souza, Marzari, and
  Vanderbilt}}]{PhysRevB.65.035109}
\bibinfo{author}{\bibfnamefont{I.}~\bibnamefont{Souza}},
  \bibinfo{author}{\bibfnamefont{N.}~\bibnamefont{Marzari}}, \bibnamefont{and}
  \bibinfo{author}{\bibfnamefont{D.}~\bibnamefont{Vanderbilt}},
  \bibinfo{journal}{Phys. Rev. B} \textbf{\bibinfo{volume}{65}},
  \bibinfo{pages}{035109} (\bibinfo{year}{2001}),
  \urlprefix\url{http://link.aps.org/doi/10.1103/PhysRevB.65.035109}.

\bibitem[{\citenamefont{Fu and Kane}(2007)}]{PhysRevB.76.045302}
\bibinfo{author}{\bibfnamefont{L.}~\bibnamefont{Fu}} \bibnamefont{and}
  \bibinfo{author}{\bibfnamefont{C.~L.} \bibnamefont{Kane}},
  \bibinfo{journal}{Phys. Rev. B} \textbf{\bibinfo{volume}{76}},
  \bibinfo{pages}{045302} (\bibinfo{year}{2007}),
  \urlprefix\url{http://link.aps.org/doi/10.1103/PhysRevB.76.045302}.

\bibitem[{\citenamefont{Fu et~al.}(2007)\citenamefont{Fu, Kane, and
  Mele}}]{PhysRevLett.98.106803}
\bibinfo{author}{\bibfnamefont{L.}~\bibnamefont{Fu}},
  \bibinfo{author}{\bibfnamefont{C.~L.} \bibnamefont{Kane}}, \bibnamefont{and}
  \bibinfo{author}{\bibfnamefont{E.~J.} \bibnamefont{Mele}},
  \bibinfo{journal}{Phys. Rev. Lett.} \textbf{\bibinfo{volume}{98}},
  \bibinfo{pages}{106803} (\bibinfo{year}{2007}),
  \urlprefix\url{http://link.aps.org/doi/10.1103/PhysRevLett.98.106803}.

\bibitem[{\citenamefont{Wang and Zhang}(2012)}]{PhysRevX.2.031008}
\bibinfo{author}{\bibfnamefont{Z.}~\bibnamefont{Wang}} \bibnamefont{and}
  \bibinfo{author}{\bibfnamefont{S.-C.} \bibnamefont{Zhang}},
  \bibinfo{journal}{Phys. Rev. X} \textbf{\bibinfo{volume}{2}},
  \bibinfo{pages}{031008} (\bibinfo{year}{2012}),
  \urlprefix\url{http://link.aps.org/doi/10.1103/PhysRevX.2.031008}.

\bibitem[{\citenamefont{Tada et~al.}(2012)\citenamefont{Tada, Peters, Oshikawa,
  Koga, Kawakami, and Fujimoto}}]{PhysRevB.85.165138}
\bibinfo{author}{\bibfnamefont{Y.}~\bibnamefont{Tada}},
  \bibinfo{author}{\bibfnamefont{R.}~\bibnamefont{Peters}},
  \bibinfo{author}{\bibfnamefont{M.}~\bibnamefont{Oshikawa}},
  \bibinfo{author}{\bibfnamefont{A.}~\bibnamefont{Koga}},
  \bibinfo{author}{\bibfnamefont{N.}~\bibnamefont{Kawakami}}, \bibnamefont{and}
  \bibinfo{author}{\bibfnamefont{S.}~\bibnamefont{Fujimoto}},
  \bibinfo{journal}{Phys. Rev. B} \textbf{\bibinfo{volume}{85}},
  \bibinfo{pages}{165138} (\bibinfo{year}{2012}),
  \urlprefix\url{http://link.aps.org/doi/10.1103/PhysRevB.85.165138}.

\bibitem[{\citenamefont{Peters et~al.}(2013)\citenamefont{Peters, Tada, and
  Kawakami}}]{PhysRevB.88.155134}
\bibinfo{author}{\bibfnamefont{R.}~\bibnamefont{Peters}},
  \bibinfo{author}{\bibfnamefont{Y.}~\bibnamefont{Tada}}, \bibnamefont{and}
  \bibinfo{author}{\bibfnamefont{N.}~\bibnamefont{Kawakami}},
  \bibinfo{journal}{Phys. Rev. B} \textbf{\bibinfo{volume}{88}},
  \bibinfo{pages}{155134} (\bibinfo{year}{2013}),
  \urlprefix\url{http://link.aps.org/doi/10.1103/PhysRevB.88.155134}.

\bibitem[{\citenamefont{Wilson}(1975)}]{wilson1975}
\bibinfo{author}{\bibfnamefont{K.~G.} \bibnamefont{Wilson}},
  \bibinfo{journal}{Rev. Mod. Phys.} \textbf{\bibinfo{volume}{47}},
  \bibinfo{pages}{773} (\bibinfo{year}{1975}),
  \urlprefix\url{http://link.aps.org/doi/10.1103/RevModPhys.47.773}.

\bibitem[{\citenamefont{Bulla et~al.}(2008)\citenamefont{Bulla, Costi, and
  Pruschke}}]{Bulla2008}
\bibinfo{author}{\bibfnamefont{R.}~\bibnamefont{Bulla}},
  \bibinfo{author}{\bibfnamefont{T.~A.} \bibnamefont{Costi}}, \bibnamefont{and}
  \bibinfo{author}{\bibfnamefont{T.}~\bibnamefont{Pruschke}},
  \bibinfo{journal}{Rev. Mod. Phys.} \textbf{\bibinfo{volume}{80}},
  \bibinfo{pages}{395} (\bibinfo{year}{2008}),
  \urlprefix\url{http://link.aps.org/doi/10.1103/RevModPhys.80.395}.

\bibitem[{\citenamefont{Peters et~al.}(2006)\citenamefont{Peters, Pruschke, and
  Anders}}]{Peters2006}
\bibinfo{author}{\bibfnamefont{R.}~\bibnamefont{Peters}},
  \bibinfo{author}{\bibfnamefont{T.}~\bibnamefont{Pruschke}}, \bibnamefont{and}
  \bibinfo{author}{\bibfnamefont{F.~B.} \bibnamefont{Anders}},
  \bibinfo{journal}{Phys. Rev. B} \textbf{\bibinfo{volume}{74}},
  \bibinfo{pages}{245114} (\bibinfo{year}{2006}),
  \urlprefix\url{http://link.aps.org/doi/10.1103/PhysRevB.74.245114}.

\bibitem[{\citenamefont{Weichselbaum and von Delft}(2007)}]{Weichselbaum2007}
\bibinfo{author}{\bibfnamefont{A.}~\bibnamefont{Weichselbaum}}
  \bibnamefont{and} \bibinfo{author}{\bibfnamefont{J.}~\bibnamefont{von
  Delft}}, \bibinfo{journal}{Phys. Rev. Lett.} \textbf{\bibinfo{volume}{99}},
  \bibinfo{pages}{076402} (\bibinfo{year}{2007}),
  \urlprefix\url{http://link.aps.org/doi/10.1103/PhysRevLett.99.076402}.

\bibitem[{\citenamefont{Rößler et~al.}(2014)\citenamefont{Rößler, Jang,
  Kim, Tjeng L.H.~Fisk, and Wirth}}]{Rossler2013}
\bibinfo{author}{\bibfnamefont{S.}~\bibnamefont{Rößler}},
  \bibinfo{author}{\bibfnamefont{T.-H.} \bibnamefont{Jang}},
  \bibinfo{author}{\bibfnamefont{D.-J.} \bibnamefont{Kim}},
  \bibinfo{author}{\bibfnamefont{Z.}~\bibnamefont{Tjeng L.H.~Fisk}},
  \bibnamefont{and} \bibinfo{author}{\bibfnamefont{S.}~\bibnamefont{Wirth}},
  \bibinfo{journal}{PNAS} \textbf{\bibinfo{volume}{111}}, \bibinfo{pages}{4798}
  (\bibinfo{year}{2014}).

\bibitem[{\citenamefont{Yee et~al.}(2013)\citenamefont{Yee, He,
  Soumyanarayanan, Kim, Fisk, and Hoffman}}]{Yee2013}
\bibinfo{author}{\bibfnamefont{M.}~\bibnamefont{Yee}},
  \bibinfo{author}{\bibfnamefont{Y.}~\bibnamefont{He}},
  \bibinfo{author}{\bibfnamefont{A.}~\bibnamefont{Soumyanarayanan}},
  \bibinfo{author}{\bibfnamefont{D.-J.} \bibnamefont{Kim}},
  \bibinfo{author}{\bibfnamefont{Z.}~\bibnamefont{Fisk}}, \bibnamefont{and}
  \bibinfo{author}{\bibfnamefont{J.}~\bibnamefont{Hoffman}}
  (\bibinfo{year}{2013}), \eprint{arXiv:1308.1085}.

\bibitem[{\citenamefont{Baruselli and Vojta}(2015)}]{Baruselli2015}
\bibinfo{author}{\bibfnamefont{P.~P.} \bibnamefont{Baruselli}}
  \bibnamefont{and} \bibinfo{author}{\bibfnamefont{M.}~\bibnamefont{Vojta}},
  \bibinfo{journal}{2D Materials} \textbf{\bibinfo{volume}{2}},
  \bibinfo{pages}{044011} (\bibinfo{year}{2015}),
  \urlprefix\url{http://stacks.iop.org/2053-1583/2/i=4/a=044011}.

\end{thebibliography}

\end{document}